\documentclass[fleqn]{pasa}

\usepackage{graphicx}
\usepackage[normal]{threeparttable}
\usepackage{gensymb}
\usepackage{mathtools, amssymb, amsthm}
\usepackage{breqn}
\usepackage{amsmath}

\newcommand\kms{km s$^{-1}$}

\newcommand\cmc{cm$^{-3}$}
\newcommand\Hii{H~\textsc{ii}}
\newcommand\emuni{pc~cm$^{-6}$}
\def \amin      {\text{$^\prime$}}

\title[ASKAP FLASH]{The First Large Absorption Survey in \mbox{H\,{\sc i}} (FLASH): I. Science Goals and Survey Design}

\author[Allison et al.]{J. R. Allison$^{1,2}$, E. M. Sadler$^{3,4,2}$, A. D. Amaral$^{5,6}$, T. An$^{7, 8}$, S. J. Curran$^{9}$, J. Darling$^{10}$, A. C. Edge$^{11}$, S. L. Ellison$^{12}$, K. L. Emig$^{13,}$\thanks{Jansky Fellow of the National Radio Astronomy Observatory}, B. M. Gaensler$^{6,5}$, L. Garratt-Smithson$^{14,2}$, M. Glowacki$^{15,16}$, K. Grasha$^{17, 2}$,  B. S. Koribalski$^{4,18}$, C. del P. Lagos$^{16,2}$, P. Lah$^{17}$, E. K. Mahony$^{4,2}$, S. A. Mao$^{19}$, R. Morganti$^{20, 21}$, V. A. Moss$^{4,3}$, M. Pettini$^{22}$, K. A. Pimbblet$^{23}$, C. Power$^{14, 2}$, P. Salas$^{25}$, L. Staveley-Smith$^{14, 2}$, M. T. Whiting$^{4}$,  O. I. Wong$^{24, 14, 2}$, H. Yoon$^{3, 2}$, Z. Zheng$^{26}$ and M. A. Zwaan$^{27}$\\

\affil{$^1$Sub-Dept. of Astrophysics, Department of Physics, University of Oxford, Denys Wilkinson Building, Keble Rd., Oxford, OX1 3RH, UK}
\affil{$^2$ARC Centre of Excellence for All-Sky Astrophysics in 3 Dimensions (ASTRO 3D)} 
\affil{$^3$Sydney Institute for Astronomy, School of Physics A28, University of Sydney, NSW 2006, Australia}%
\affil{$^4$ATNF, CSIRO, Space and Astronomy, Australia Telescope National Facility, PO Box 76, Epping, NSW 1710, Australia}
\affil{$^5$David A. Dunlap Department of Astronomy and Astrophysics, University of Toronto, ON, M5S 3H4, Canada}
\affil{$^6$Dunlap Institute for Astronomy and Astrophysics, University of Toronto, Toronto, ON, M5S 3H4, Canada}
\affil{$^{7}$Shanghai Astronomical Observatory, Chinese Academy of Sciences, Nandan Road 80, Shanghai 200030, China}
\affil{$^{8}$Key Laboratory of Radio Astronomy, Chinese Academy of Sciences, Nanjing 210008, China}
\affil{$^{9}$School of Chemical and Physical Sciences, Victoria University of Wellington, PO Box 600, Wellington 6140, New Zealand}
\affil{$^{10}$CASA, Department of Astrophysical and Planetary Sciences, University of Colorado, 389 UCB,
Boulder, CO 80309-0389, USA}
\affil{$^{11}$Centre for Extragalactic Astronomy, Durham University, Durham, DH1 3LE, UK}
\affil{$^{12}$Department of Physics \& Astronomy, University of Victoria, Finnerty Road, Victoria, British Columbia, V8P 1A1, Canada}
\affil{$^{13}$National Radio Astronomy Observatory, 520 Edgemont Road, Charlottesville, VA 22903, USA}
\affil{$^{14}$ICRAR, University of Western Australia, 35 Stirling Highway, Crawley, Western Australia 6009, Australia}
\affil{$^{15}$ICRAR, Curtin University, Bentley, WA 6102, Australia}
\affil{$^{16}$Inter-University Institute for Data Intensive Astronomy, Bellville 7535, South Africa}
\affil{$^{17}$Research School of Astronomy and Astrophysics, Australian National University, Canberra, ACT 2611, Australia}
\affil{$^{18}$Western Sydney University, Locked Bag 1797, Penrith, NSW 2751, Australia}
\affil{$^{19}$Max Planck Institute for Radio Astronomy, Auf dem H\"ugel 69, Bonn D-53121, Germany}
\affil{$^{20}$ASTRON, the Netherlands Institute for Radio Astronomy, Oude Hoogeveensedijk 4, 7991 PD Dwingeloo, The Netherlands}
\affil{$^{21}$Kapteyn Astronomical Institute, University of Groningen, Postbus 800, 9700 AV Groningen, The Netherlands}
\affil{$^{22}$Institute of Astronomy, University of Cambridge, Madingley Road, Cambridge, CB3 0HA, UK}
\affil{$^{23}$E.A.Milne Centre for Astrophysics, University of Hull, Cottingham Road, Kingston-upon-Hull, HU6 7RX, UK}
\affil{$^{24}$ATNF, CSIRO, Space and Astronomy, PO Box 1130, Bentley WA 6102, Australia}
\affil{$^{25}$Green Bank Observatory, 155 Observatory Road, Green Bank, WV 24915, USA}
\affil{$^{26}$National Astronomical Observatories, Chinese Academy of Sciences, 20A Datun Road, Chaoyang District, Beijing 100101, China}
\affil{$^{27}$European Southern Observatory, Karl-Schwarzschild-Str. 2, 85748 Garching b. M\"unchen, Germany}
}%

\jid{PASA}
\doi{10.1017/pas.\the\year.xxx}
\jyear{\the\year}

\usepackage{aas_macros}
\usepackage{hyperref} 
\hypersetup{colorlinks,citecolor=blue,linkcolor=blue,urlcolor=blue}

\begin{document}

\begin{frontmatter}
\maketitle

\begin{abstract}
We describe the scientific goals and survey design of the First Large Absorption Survey in \mbox{H\,{\sc i}} (FLASH), a wide field survey for 21-cm line absorption in neutral atomic hydrogen (\mbox{H\,{\sc i}}) at intermediate cosmological redshifts. FLASH will be carried out with the Australian Square Kilometre Array Pathfinder (ASKAP) radio telescope and is planned to cover the sky south of $\delta \approx +40\,\deg$ at frequencies between 711.5 and 999.5\,MHz. At redshifts between $z = 0.4$ and $1.0$ (look back times of 4 -- 8\,Gyr), the \mbox{H\,{\sc i}} content of the Universe has been poorly explored due to the difficulty of carrying out radio surveys for faint 21-cm line emission and, at ultra-violet wavelengths, space-borne searches for Damped Lyman-$\alpha$ absorption in quasar spectra. The ASKAP wide field of view and large spectral bandwidth, in combination with a radio-quiet site, will enable a search for absorption lines in the radio spectra of bright continuum sources over 80\% of the sky. This survey is expected to detect at least several hundred intervening 21-cm absorbers, and will produce an \mbox{H\,{\sc i}}-absorption-selected catalogue of galaxies rich in cool, star-forming gas, some of which may be concealed from optical surveys. Likewise, at least several hundred associated 21-cm absorbers are expected to be detected within the host galaxies of radio sources at $0.4 < z < 1.0$, providing valuable kinematical information for models of gas accretion and jet-driven feedback in radio-loud active galactic nuclei. FLASH will also detect OH 18-cm absorbers in diffuse molecular gas, megamaser OH emission, radio recombination lines, and stacked \mbox{H\,{\sc i}} emission.\\
\end{abstract}

\end{frontmatter}


\section{Introduction}
\label{sec:intro}

Hydrogen is the most abundant element of baryonic matter in the Universe. The neutral atomic phase (\mbox{H\,{\sc i}}) is prevalent throughout the interstellar medium (ISM) and is observable through the 21-cm (1420.4\,MHz hyperfine splitting of the $n=1$ ground state) and Lyman-$\alpha$ ($\lambda = 1215.6$\,\AA; $n = 2 - 1$) transitions. It is therefore an important tracer of the abundance and kinematics of the neutral gas in galaxies throughout cosmic history. However, sensitivity-limited radio surveys for the 21-cm emission line have so far only detected \mbox{H\,{\sc i}} in individual galaxies at cosmological redshifts up to $z \sim 0.3$ (e.g. \citealt{Catinella:2015, Fernandez:2016}). 

Further study using the 21-cm emission line beyond the low-$z$ Universe requires a statistical approach either by averaging the spectra of known galaxies (e.g. \citealt{Rhee:2018, Bera:2019, Chowdhury:2020a}) or using intensity mapping (e.g. \citealt{Chang:2010, Masui:2013}). New \mbox{H\,{\sc i}} surveys with the pathfinder and precursor telescopes to the Square Kilometre Array (SKA) are expected to greatly increase the cosmological volume in which 21-cm emission is detected (e.g. \citealt{Blyth:2016, Adams:2019, Koribalski:2020, Maddox:2021}). Yet despite significant improvements in the collecting areas, fields-of-view and available frequency ranges of these radio telescopes, it is still significantly challenging to detect 21-cm emission from individual galaxies at redshifts much beyond the local Universe ($z \gg 0.1$).

An alternative approach is to detect \mbox{H\,{\sc i}} absorption lines in the spectra of background continuum sources, for which the sensitivity is independent of luminosity distance and hence redshift. Indeed almost all current information about the \mbox{H\,{\sc i}} in the high-redshift Universe has been obtained from surveys of Lyman-$\alpha$ absorption in the spectra of quasars (e.g. \citealt{Noterdaeme:2012, Bird:2017}). However, at redshifts below $z \sim 1.7$ the Lyman-$\alpha$ line shifts from the optical to the ultra-violet (UV) part of the spectrum where it is only observable with space-borne telescopes (e.g. \citealt{Neeleman:2016, Rao:2017}). This has led to a situation where our knowledge of the \mbox{H\,{\sc i}} content of the Universe between $z \sim 0.5$ and 2 is less complete than at higher redshifts.

In this paper we describe the parameters and scientific motivation for a radio survey that is designed to search for \mbox{H\,{\sc i}} in galaxies at these poorly-explored redshifts. The First Large Absorption Survey in \mbox{H\,{\sc i}} (FLASH) will use the Australian Square Kilometre Array Pathfinder (ASKAP; \citealt{DeBoer2009, Hotan:2021}) to detect 21-cm absorption lines in the continuum spectra of radio sources over 80\% of the sky. ASKAP is one of the key precursor instruments of the Square Kilometre Array (SKA) and implements pioneering phased-array-feed (PAF) receiver technology that increases the field of view to 30 times that expected from a standard receiver (\citealt{Hay:2008}). 

The available ASKAP frequency range from 700 to 1800\,MHz allows observations of the H{\sc i} 21-cm line over a wide range in redshift from the local Universe to $z \sim 1$. This will enable transformational science to be carried out by detecting H{\sc i} in large numbers of distant galaxies, as discussed in the original ASKAP science case \citep{Johnston2008}. Both the field of view and wide bandwidth of this interferometer enables searches for H{\sc i} absorption to be carried out quickly and efficiently. Of particular importance is the radio quietness of the ASKAP site, particularly at frequencies between 700 and 1000\,MHz corresponding to uninterrupted \mbox{H\,{\sc i}} redshift coverage between $z = 0.4$ and $1.0$ (see e.g. \citealt{Allison:2017}). 

Between 2014 and 2016, the first six dishes of ASKAP were fitted with Mark I PAFs and operated as the Boolardy Engineering Test Array (BETA; \citealt{McConnell2016}). \cite{Allison:2015} used BETA to obtain a spectroscopically-blind detection of associated \mbox{H\,{\sc i}} absorption towards the GHz-peaked spectrum radio galaxy PKS\,B1740$-$517. Further work using BETA confirmed that ASKAP could be used to routinely detect \mbox{H\,{\sc i}} and OH absorption towards a range of sources at the redshifts required by FLASH (\citealt{Allison:2016a, Moss:2017, Allison:2017}). 

At the end of 2016, a sub-array of twelve dishes (ASKAP-12) was fitted with new Mark II PAFs  that have improved performance over the majority of the frequency band (e.g. \citealt{Chippendale:2015}). A further paper on PKS\,B1740$-$517 by \cite{Allison:2019} demonstrated the continued capability of ASKAP to detect  \mbox{H\,{\sc i}} absorption using these Mark II PAFs. Work by \cite{Glowacki:2019} and \cite{Sadler:2020} also demonstrated the continued success of using ASKAP to detect \mbox{H\,{\sc i}} absorption towards larger samples of targeted radio sources. 

Using ASKAP-12, \cite{Allison:2020} carried out an early-science FLASH survey of the GAMA\,23 field (\citealt{Liske:2015}) to search for \mbox{H\,{\sc i}} absorption over a sky area of approximately 50 square degrees. This work demonstrated the feasibility of detecting extragalactic absorption lines in a completely un-targeted wide-field radio survey. 

The full 36-antenna ASKAP array is now available and FLASH has recently completed Phase I (the first 100\,hrs) of its pilot survey, covering approximately 1000\,$\deg^{2}$ over several large fields in the southern hemisphere and equator. While this paper focuses on the science cases and design for FLASH, an accompanying paper (Yoon et al., in preparation) will present an overview of the data products and early results from the pilot survey. 

We structure this paper as follows: in \autoref{section:hi-absorption_in_galaxies} we discuss the background and current status of 21-cm absorption surveys, in \autoref{section:survey_description} we describe the survey parameters and present the main science goals in \autoref{section:science_goals}, in \autoref{section:simulations} we discuss the links to theoretical models simulations, a description of the expected data products is given in \autoref{section:data_products}, and a brief summary of the survey can be found in \autoref{section:summary}. The interested reader can find further details about how the survey outcomes are calculated in \autoref{section:estimating_survey_outcomes}. Where relevant we use a flat $\Lambda$-CDM cosmology with $\Omega_{\rm m} = 0.3$, $\Omega_{\Lambda} = 0.7$ and $H_{0} = 70$\,km\,s$^{-1}$\,Mpc$^{-1}$.


\section{\mbox{H\,{\sc i}} 21-cm absorption in galaxies}\label{section:hi-absorption_in_galaxies}

The 21-cm absorption line of \mbox{H\,{\sc i}}, detected in the spectra of background radio sources, is a useful tool for measuring the cold gas content of galaxies at large cosmological redshifts. The absorption line flux density is independent of the luminosity distance to the absorber, and so for a sufficiently bright sample of background radio sources, enables surveys of neutral gas at distances well beyond that currently obtainable with emission line surveys. 

\subsection{Physical and observable properties}


\subsubsection{Optical depth and column density}

The main observable in any absorption line survey is the ratio of the line ($\Delta{S}$) to continuum flux density ($S_{\rm c}$), which is used to determine the optical depth $\tau$ as a function of velocity $v$ as follows
\begin{equation}\label{equation:optical_depth}
	\tau(v) = -\ln{\left(1 + \frac{\Delta{S}(v)}{c_{\rm f}\,S_{\rm c}(v)}\right)}.
\end{equation}
The source covering factor, $c_{\rm f}$, accounts for the areal fraction of continuum flux density that is subtended by the absorber of average optical depth $\tau$. For a transition ${\rm l} \rightarrow {\rm u}$ of species X, the column density of particles inferred from the velocity-integrated optical depth, $\int{\tau_{\rm lu} (v)\,\mathrm{d}{v}}$, is then given by
\begin{equation}
N_{\rm X} = \frac{8\pi}{c^{3}}\frac{\nu_{\rm lu}^{3}}{g_{\rm u}}\frac{f(T_{\rm lu})}{A_{\rm ul}}\int{\tau_{\rm lu} (v)\,\mathrm{d}{v}},
\end{equation}
where 
\begin{equation}
f(T_{\rm lu}) = \frac{Q(T_{\rm lu})\,e^{E_{u}/k_{\rm B}T_{\rm lu}}}{e^{h\nu_{\rm lu}/k_{\rm B}T_{\rm lu}}-1},	
\end{equation}
and $\nu_{\rm lu}$ is the rest frequency, $g_{\rm u}$ is the statistical weight of $u$, $A_{\rm ul}$ is the Einstein coefficient for spontaneous emission, $E_{u}$ is the upper energy level, and $Q(T_{\rm lu})$ is the partition function assuming a single excitation temperature, $T_{\rm lu}$ (e.g. \citealt{Wiklind:1995}). 

For the \mbox{H\,{\sc i}} 21-cm line we ignore $n > 1$ states, so that $Q(T_{\rm lu}) = g_{\rm l}\exp(-E_{\rm l}/k_{\rm B}T_{\rm lu}) + g_{\rm u}\exp(-E_{\rm u}/k_{\rm B}T_{\rm lu})$, and $h\,\nu_{\rm lu} \ll k_{\mathrm B}\,T_{\rm lu}$, so that
\begin{equation}
	f(T_{\rm lu}) = \frac{g_{\rm l}\,e^{h\nu_{\rm lu}/k_{\rm B}T_{\rm lu}} + g_u}{e^{h\nu_{\rm lu}/k_{\rm B}T_{\rm lu}}-1} \approx (g_{\rm l} + g_{\rm u})\frac{k_{\rm B}T_{\rm lu}}{h\nu_{\rm lu}}, 
\end{equation}
where $g_{\rm l} = 1$ and $g_{\rm u} = 3$. Denoting the 21-cm excitation temperature as $T_{\rm lu} = T_{\rm s}$ (the spin temperature) and using the known Einstein coefficient $A_{\rm ul} = 2.85 \times 10^{-15}$\,s$^{-1}$, the \mbox{H\,{\sc i}} column density (in atoms\,cm$^{-2}$) is then given by 
\begin{equation}\label{equation:column_density}
    N_{\rm HI} \approx 1.823 \times 10^{18}~T_{\rm s}~\int{\tau(v)\,\mathrm{d}v}, 
\end{equation}   
where $T_{\rm s}$ is in K and the velocity $v$ is in km\,s$^{-1}$. This is the key equation that relates the inferred 21-cm optical depth to the physical properties of the \mbox{H\,{\sc i}} gas.


\subsubsection{The spin temperature}

The relationship given by \autoref{equation:column_density} shows that for a fixed column density of neutral gas the 21-cm optical depth is inversely proportional to the spin temperature. Depending on the gas density and intensity of illuminating sources, excitation of the hyperfine transition in \mbox{H\,{\sc i}}, and hence $T_{\rm s}$, is determined by particle collisions, Lyman-$\alpha$ radiation, or 21-cm radiation (e.g. \citealt{Wouthuysen:1952, Purcell:1956, Field:1958, Field:1959, Bahcall:1969, Liszt:2001}). 

In our own Galaxy, where multiple sight lines with simultaneous 21-cm emission and absorption enable detailed study of the physical conditions, the neutral ISM is found to comprise two stable thermal phases in pressure equilibrium; the cold (CNM) and warm (WNM) neutral medium (\citealt{Wolfire:2003}). There is also observational evidence for a significant fraction of  an intermediate unstable phase (UNM) that is thought to be generated by dynamical processes such as turbulence and supernova shocks (e.g. \citealt{Heiles:2003, Murray:2018}). 

In the denser CNM gas ($n \sim 100$\,cm$^{-3}$), particle collisions thermalise the 21-cm transition so that the spin temperature is approximately equal to the kinetic temperature ($T_{\rm s} \approx T_{\rm k} \sim 100$\,K). Although this is not the case in the more diffuse WNM gas ($n \sim 0.1$\,cm$^{-3}$), excitation of the 21-cm transition is still coupled to the kinetic temperature of the gas through a combination of particle collisions and scattering of the ambient Lyman-$\alpha$ radiation field (the Wouthuysen-Field effect; \citealt{Wouthuysen:1952, Field:1958, Field:1959}), so that $T_{\rm s} \lesssim T_{\rm k} \sim 10\,000$\,K (\citealt{Liszt:2001}; \citealt{Murray:2018}). 

Hence, for typical ISM conditions, the spin temperature is an increasing function of the gas kinetic temperature so that any measurement of the absorption line is weighted in favour of colder \mbox{H\,{\sc i}} gas on our line of sight. The relationship in \autoref{equation:column_density} can therefore either be thought of as the inferred column density of total \mbox{H\,{\sc i}} gas, where $T_{\rm s}$ is the $N_{\rm HI}$-weighted harmonic mean over all thermal components, or as the column density of cooler absorbing \mbox{H\,{\sc i}} gas, in which case $T_{\rm s}$ is the spin temperature of the absorbing component. 

In the Milky Way ISM the neutral phases exist in the mass ratio CNM:UNM:WNM = 28:20:52 (\citealt{Murray:2018}), corresponding to a mass-weighted harmonic mean spin temperature of $T_{\rm s} \approx 300$\,K (see also \citealt{Dickey:2009}). At greater distances, $T_{\rm s}$ can be measured for random sight lines through galaxies where an existing measurement of the \mbox{H\,{\sc i}} column density is available from 21-cm emission or Lyman-$\alpha$ absorption. The $T_{\rm s}$ measured for Damped Lyman-$\alpha$ Absorbers (DLAs; $N_{\rm HI} \geq 2 \times 10^{20}$\,cm$^{-2}$) varies significantly between $\approx 100$ and $10\,000$\,K, presumably due to variance in the physical conditions of the gas probed by each sight line (see \citealt{Kanekar:2014a}). However, despite this sight-line variance, the harmonic mean $T_{\rm s}$ for all DLAs at $z < 1$ is consistent with that of the Milky Way ISM (\citealt{Allison:2021}). We therefore adopt 300\,K as our fiducial value of the spin temperature here.

Since the 21-cm absorption line is sensitive to the colder \mbox{H\,{\sc i}} gas, it is complementary to the 21-cm emission and Lyman-$\alpha$ absorption lines that trace the total \mbox{H\,{\sc i}} gas, and is an important probe of the evolution of star forming gas in galaxies over the history of the Universe.


\subsection{Current status of observational work}\label{section:current_status_observations}

Since the first detections of the 21-cm absorption line in NGC\,5128/Centaurus A (\citealt{Roberts:1970}) and M\,82 (\citealt{Guelin:1970}), this transition has been used to study the abundance and kinematics of extragalactic cold \mbox{H\,{\sc i}} gas. Detections are broadly classified as either intervening or associated/intrinsic, depending on whether they arise in a separate (on cosmological scales) foreground galaxy, or in the host galaxy of the radio source itself. Intervening 21-cm absorbers are used to carry out a census of the cold gas in the Universe, analogous to the Damped Lyman-$\alpha$ Absorber (DLA) surveys in the UV and visible bands (e.g. \citealt{Noterdaeme:2012, Zafar:2013, Crighton:2015, Sanchez-Ramirez:2016, Neeleman:2016, Bird:2017, Rao:2017}). Separately, associated absorbers are widely used to determine the kinematics of neutral gas associated with active galaxies and mechanisms for accretion and feedback (see \citealt{Morganti:2018} for a detailed review of this subject).


\subsubsection{Intervening 21-cm absorbers}

Until recently, surveys for intervening 21-cm absorbers have been largely constrained by narrow bandwidths and/or fields of view to targeted observations of the following: 
\begin{itemize}
    \item 
Known `quasar-galaxy pairs', where the line of sight to a distant radio quasar passes close to a galaxy in the nearby Universe (e.g. \citealt{Gupta:2010, Borthakur:2011, Reeves:2015, Reeves:2016, Borthakur:2016, Dutta:2017a}), 
\item 
Sight lines with known optical/UV DLAs (see \citealt{Kanekar:2014a} and references therein), 
\item 
Known (optical) Mg\,{\sc ii}\,$\lambda\lambda$\,2796, 2803\,\AA\ absorbers, used as a proxy for DLAs at $z<1.7$\ (e.g. \citealt{Lane:2000, Lane:2001, Kanekar:2009a, Gupta:2009, Gupta:2012, Dutta:2020}), and 
\item{\mbox{Fe\,{\sc ii}}}\,$\lambda$2600\,\AA\ absorbers (e.g. \citealt{Dutta:2017b}). 
\end{itemize}
Detection rates among quasar-galaxy pairs in the nearby Universe are typically 5 -- 15\,\% and are inversely related to the impact parameter, suggesting that the coldest \mbox{H\,{\sc i}} gas detected in 21-cm absorption is largely confined to the inner discs of galaxies (e.g. \citealt{Borthakur:2016, Curran:2016b, Dutta:2017a, Curran:2020}). 

The detection rates for 21-cm absorbers in redshifted DLAs are typically higher ($\sim 50$\,\%; \citealt{Kanekar:2014a}) and consistent with sight lines that are known to intercept high-column-density neutral gas. 21-cm line observations of DLAs provide some evidence for an increase in the mean spin temperature at high redshifts, and an anti-correlation with metallicity, that would be expected for an evolution of the neutral gas that follows the star formation history of the Universe (\citealt{Kanekar:2014a, Allison:2021}, see also \citealt{Curran:2019a}). 


\subsubsection{Associated 21-cm absorbers}\label{section:observations_associated_absorbers}

The majority of searches for associated 21-cm absorbers have been limited to radio sources with known optical spectroscopic redshifts. It is possible that this may select against detection of \mbox{H\,\sc i} at high redshifts ($z \gtrsim 1$), since optically identified active galactic nuclei (AGN) that are UV luminous in their rest-frame could ionize or excite the neutral gas (\citealt{Curran:2008, Curran:2010, Curran:2012}). This hypothesis is supported by observational evidence that the absorption signal strength is inversely related to the UV luminosity, and that very few 21-cm absorbers are detected in AGN for $L_{\rm UV} \gtrsim 10^{23}$\,W\,Hz$^{-1}$ (e.g. \citealt{Curran:2013, Aditya:2018b, Curran:2019b, Grasha:2019, Chowdhury:2020b, Mhaskey:2020}, see also \citealt{Aditya:2021}). 

Likewise, since 21-cm photons directly excite the hyperfine transition, a similar selection bias may also exist for radio luminosity (e.g. \citealt{Curran:2008, Aditya:2016, Aditya:2018b}). However, as yet there is no evidence that this affects the detection rate of associated 21-cm absorbers (e.g. \citealt{Curran:2019b, Grasha:2019}). 

The factors that determine the detection of \mbox{H\,{\sc i}} absorption in active galaxies are complex and require careful consideration of the properties of the radio sources and their hosts. Reasonably high detection rates ($\sim 30$\,\%) are achieved towards compact steep spectrum (CSS) and GHz-peaked spectrum (GPS) radio sources, of which those that are associated with radio galaxies, rather than quasars (i.e. type-1 AGN), are most prolific (e.g. \citealt{Veron-Cetty:2000, Vermuelen:2003, Gupta:2006, Chandola:2011, Gereb:2015, Aditya:2018a}). These are either young ($t_{\rm age} \lesssim 10$\,kyr) or confined older radio sources \citep{An:2012}, with linear extents typically smaller than their host galaxies ($d \lesssim 10$\,kpc) and so preferentially located behind high column densities of neutral gas, possibly in a circumnuclear disc or torus (\citealt{Pihlstrom:2003, Orienti:2006, Curran:2013}). 

Similarly, searches for 21-cm absorption in highly dust-reddened quasars, interacting, and merging galaxies have been very successful in detecting associated absorbers ($\sim$\,80\,\%; e.g. \citealt{Carilli:1998, Yan:2016, Maccagni:2017, Dutta:2018, Dutta:2019}), again consistent with a model whereby accreted circumnuclear gas has not yet been cleared away by the active nucleus. In contrast, larger radio sources have a significantly lower detection rate ($\sim 15$\,\%; e.g. \citealt{Morganti:2001, Gupta:2006, Chandola:2013, Maccagni:2017}), where a large fraction of the flux density is located beyond the extent of \mbox{H\,{\sc i}} gas in the host galaxy. In all cases these previous surveys were targeted towards sources that were selected based on their core flux density, leading to a selection bias that favours detection of associated absorption. Hence although future wide-field 21-cm absorption line surveys are expected to yield far more associated absorbers at cosmological distances than previously achieved, the detection rates are likely to be significantly lower (e.g. \citealt{Allison:2014, Allison:2020}).

\begin{figure}
\begin{center}
\includegraphics[width=\columnwidth]{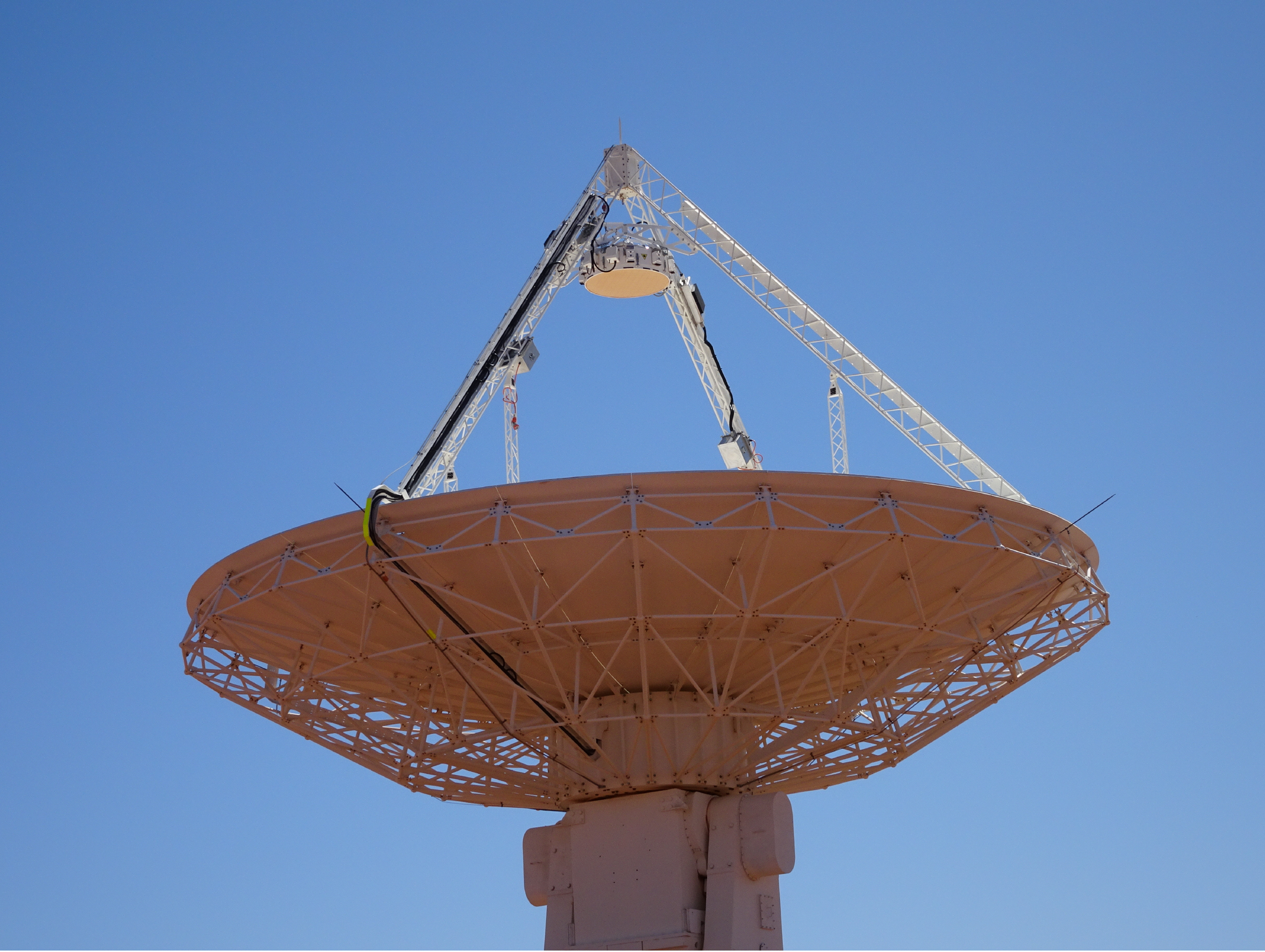}
\caption{A 12-m ASKAP antenna, equipped with a Mark-II Phased Array Feed -- Photo Credit: Robert Hollow, CSIRO.}\label{figure:askap_antenna}
\end{center}
\end{figure}


\subsubsection{Wide-field \& spectroscopically blind surveys}

As already discussed, previous work has focused mainly on targeted searches for extragalactic 21-cm absorption lines and as yet there have been few large-area, spectroscopically-blind surveys. Notably, in early work \cite{Darling:2004} detected a 21-cm absorber from a large spectroscopically blind survey of radio sources using the Green Bank Telescope (GBT). Later, \cite{Darling:2011} carried out the first wide-field 21-cm absorption in the nearby Universe, using early data from the Arecibo Legacy Fast Arecibo L-band Feed Array (ALFALFA) survey (\citealt{Giovanelli:2005}), searching for \mbox{H\,{\sc i}} absorption against more than 7000 radio sources at $z < 0.058$ over a 517\,deg$^2$ area of sky. Although they only detected a single associated absorber (in the merging system UGC\,6081), their results within this volume were consistent with that expected given the known \mbox{H\,{\sc i}} column density frequency distribution function and for spin temperatures greater than 100\,K. More recently, \cite{Grasha:2020} used the GBT to carry out a spectroscopically blind survey of 252 radio sources covering redshifts between $z = 0$ and $2.74$, obtaining ten detections of known absorbers. Again, their results were consistent with expectations based on the known distribution of \mbox{H\,{\sc i}} column densities in the local and high-$z$ Universe, and spin temperatures in the typical ISM range.

The new surveys will build on this earlier, single-dish work, making use of advances in radio telescope technology and radio-quiet observatories to detect 21-cm absorption across a range of redshifts towards thousands of radio sources. Notably the new interferometers, such as ASKAP and the South African Meer-Karoo Array Telescope (MeerKAT; \citealt{Jonas:2018}), are particularly well suited to this task since they can provide sufficiently flat spectral baselines across the bandwidths required to cover these larger redshift intervals (e.g. \citealt{Allison:2020, Gupta:2021}). 


\section{Survey description}\label{section:survey_description}

The principal goal of FLASH is to detect extragalactic \mbox{H\,{\sc i}} 21-cm absorption lines towards radio sources in the southern sky. 
It is one of the eight Survey Science Projects (SSPs) to be undertaken with ASKAP. In this section we describe the important parameters and compare them with similar surveys on other telescopes. 


\subsection{The Australian SKA Pathfinder telescope }

\subsubsection{Overview}

ASKAP is a radio interferometer that comprises 36 identical 12\,m antennas, each equipped with a PAF that can be used to electronically form up to 36 beams that sample a 31\,deg$^{2}$ field of view at 800\,MHz (\citealt{Hotan:2021}; see \autoref{figure:askap_antenna}). It is located at the Murchison Radio Astronomy Observatory (MRO), near the Boolardy Station in Murchison Shire, Western Australia (26\degree42$^{\prime}$11$^{\prime\prime}$ S, 116\degree40$^{\prime}$14$^{\prime\prime}$ E). The telescope operates at frequencies between 700 and 1800\,MHz and can therefore be used to detect the \mbox{H\,{\sc i}} 21-cm line at redshifts up to $z = 1$, and the OH 18-cm lines up to $z = 1.37$. 

When making full use of the available 288\,MHz bandwidth\footnote{Future upgrades to the ASKAP correlator may include an increase in bandwidth to 336\,MHz.} the correlator generates 18.5\,kHz channels, corresponding to radial velocities of $3.1$\,km\,s$^{-1}$ at 1800\,MHz and $7.9$\,km\,s$^{-1}$ at 700\,MHz. This spectral resolution is well matched to the expected line widths of intervening 21-cm absorbers, with at least three channels across the average FWHM of lines reported in the literature ($\approx 30$\,km\,s$^{-1}$; see \citealt{Curran:2016a}).

The array consists of a ``core'' of 30 antennas that give good surface brightness sensitivity, and six outer antennas that provide up to 6\,km baselines for higher spatial resolution imaging. In the case of FLASH, the continuum imaging will make full use of the longest antenna baselines, with a spatial resolution of about $10$\,arcsec, to provide as much source morphology information as possible. The spectral-line cubes will be imaged with close-to-natural weighting ($\approx 20$\,arcsec) to retain optimal sensitivity for absorption-line detection. We note that this resolution is not sufficient to unambiguously determine the spatial distribution of detected absorbers and/or background sources, the majority of which will be unresolved (e.g. \citealt{Becker:1995}). Further spatial interpretation of the FLASH-detected absorbers and their background sources will require follow-up observations at $\sim 10$\,mas resolution using very long baseline interferometry (e.g. \citealt{Braun:2012}). 

\subsubsection{ASKAP sensitivity}

The sensitivity of a radio telescope is typically defined in terms of its system temperature ($T_{\rm sys}$), antenna efficiency ($\eta$) and collecting area ($A$). For ASKAP, the ratio $T_{\rm sys}/\eta$ varies from a maximum of about 120\,K at 700\,MHz to 65\,K at 1300\,MHz (\citealt{Hotan:2021}), which for the single antenna area of $A = 113.1$\,m$^{2}$ correspond to system equivalent flux densities (SEFD, $S_{\rm sys}$) between 2930 and 1590\,Jy. The expected noise per spectral channel ($\sigma_{\rm chan}$) can then be calculated using the following radiometer equation for interferometric arrays,
\begin{equation}
    \sigma_{\rm chan} = \frac{S_{\rm sys}}{\sqrt{n_{\rm pol}\,n_{\rm ant}(n_{\rm ant}-1)\,\Delta{t}\,\Delta{\nu_{\rm chan}}}},
\end{equation}
where $n_{\rm pol}$ is the number of polarization channels, $n_{\rm ant}$ is the number of antennas, $\Delta{t}$ is the integration time and $\Delta{\nu_{\rm chan}}$ is the channel bandwidth. Within a 2\,hour observation with the 36-antenna dual-polarization array the noise level is expected to be between 2.7 and 5.1\,mJy\,beam$^{-1}$ per 18.5\,kHz channel across the full range of available ASKAP frequencies\footnote{This assumes that the spectral-line data cubes are imaged with close-to-natural weighting.}.


\subsection{Survey design and parameters}

FLASH will be a rapid survey of the southern sky that optimises the use of the wide-field of view and quiet RFI environment of ASKAP for detection of 21-cm absorption. We summarise the important survey parameters in \autoref{table:survey_parameters}.

\begin{table}
\caption{Summary of the key FLASH survey parameters.}
\centering
\def\arraystretch{1.1}
\begin{threeparttable}
\begin{tabular}{@{}lccc@{}}
\hline\hline 
No. pointings & \multicolumn{3}{c}{903} \\
Total sky area & \multicolumn{3}{c}{34\,000\,$\deg^{2}$} \\
Time per pointing & \multicolumn{3}{c}{2\,h} \\
Total survey time & \multicolumn{3}{c}{1806\,h} \\
$\nu_{\rm obs}$ & \multicolumn{3}{c}{711.5 -- 999.5\,MHz} \\
$z_{\rm HI}$ & \multicolumn{3}{c}{0.4 -- 1.0} \\
$\Delta{\nu_{\rm chan}}$ & \multicolumn{3}{c}{18.5\,kHz} \\
$\Delta{v_{\rm chan}}$ & \multicolumn{3}{c}{5.5 -- 7.8\,km\,s$^{-1}$} \\
$\sigma_{\rm chan}$\tnote{(a)} & \multicolumn{3}{c}{3.2 -- 5.1\,mJy\,beam$^{-1}$} \\
\hline\hline
\end{tabular}
\begin{tablenotes}
\item[a] Based on the SEFD measured by \citet{Hotan:2021} and natural weighting.
\end{tablenotes}
\end{threeparttable}
\label{table:survey_parameters}
\end{table}


\subsubsection{Choice of survey frequency and sky coverage}

The baseline survey parameters of FLASH are 2\,h per pointing covering the entire southern sky below $\delta \approx +40\,\deg$, at frequencies between 711.5 and 999.5\,MHz. These were chosen to optimise the discovery potential and detection yield for 21-cm absorption, which is quantified in terms of the co-moving absorption path length ($\Delta{X}$). This is the total co-moving interval over which intervening 21-cm absorbers may be detected, thus also providing a metric by which other 21-cm line surveys can be compared with FLASH.

To estimate $\Delta{X}$ we use the completeness-corrected procedure described in \autoref{section:estimating_survey_outcomes}. This takes into account the completeness to spectral lines of a specific width and peak signal-to-noise, based on the analysis of an ASKAP-12 survey by \cite{Allison:2020}. The data from that earlier survey contained a channelisation error that contributed a multiplicative non-Gaussian component to the noise level, thus elevating the threshold required to achieve a reliable detection. Although this channelisation error has now been corrected it is likely that any blind absorption-line survey such as FLASH will still include a background level of false-positive detections due to multiplicative non-Gaussian noise (resulting from either hardware and/or data processing errors). We therefore consider the earlier ASKAP-12 data to be representative of FLASH, but expect that the false-positive error rate may decrease over time as our identification techniques improve.

To simulate a realistic wide-field survey with ASKAP, we use an input catalogue of background sources using the National Radio Astronomy Observatory Very Large Array Sky Survey (NVSS; \citealt{Condon:1998}) catalogue, and apply a statistical distribution for the source redshifts based on the model of \cite{deZotti:2010} (see \autoref{section:estimating_survey_outcomes} for further details). 

\begin{figure}
\begin{center}
\includegraphics[width=\columnwidth]{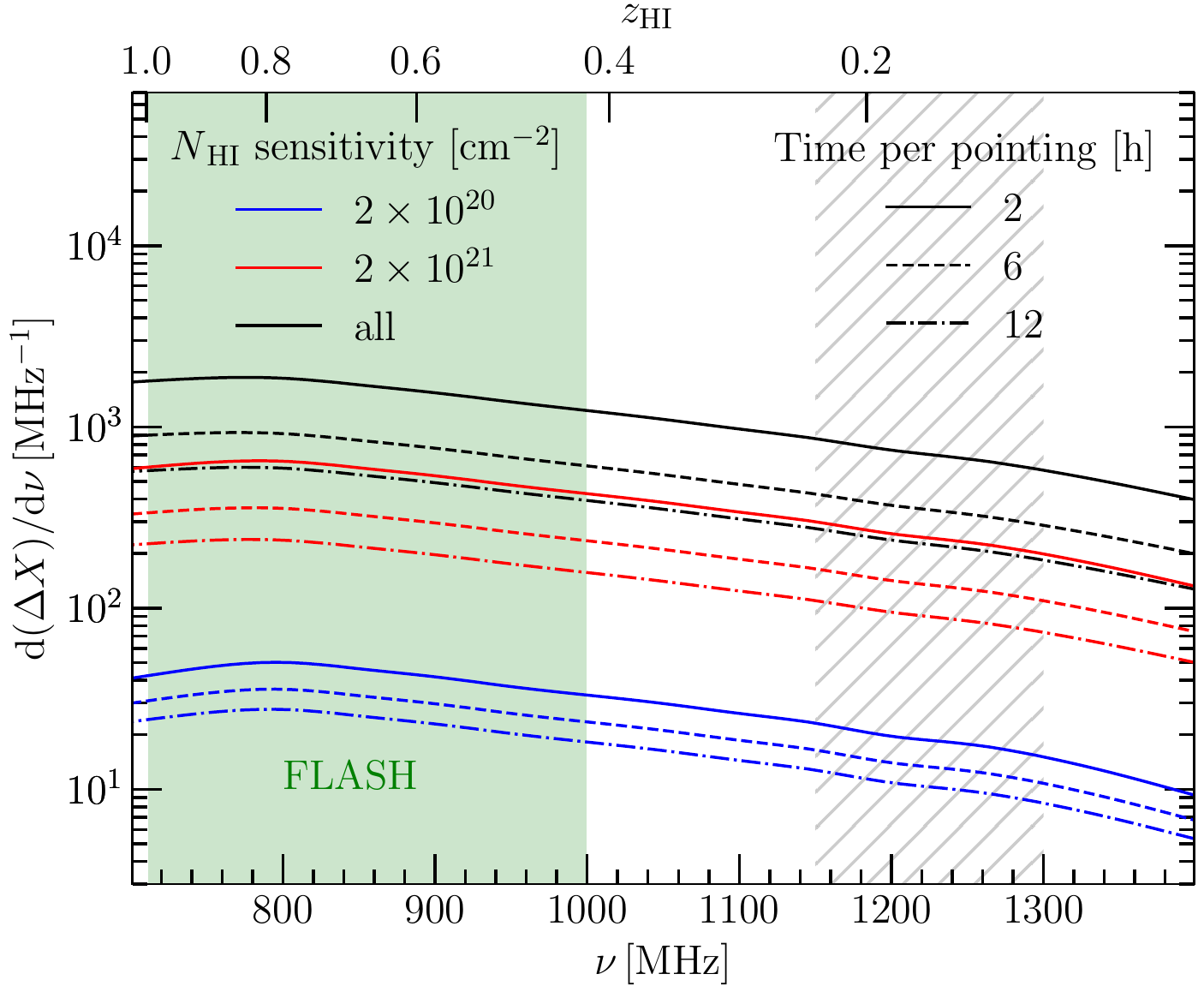}
\caption{The 21-cm absorption path length  ($\Delta{X}$) as a function of observed frequency ($\nu$) across the ASKAP band. Solid lines denote the baseline FLASH survey parameters, which are a 2\,h integration time per pointing, covering the entire sky south of $\delta \approx +40\deg$. Dashed and dot-dashed lines correspond to higher integration times per pointing for the same total survey time. Coloured lines represent column density sensitivity limits of $N_{\rm HI} = 2 \times 10^{20}$ (blue) and $N_{\rm HI} = 2 \times 10^{21}$\,cm$^{-2}$ (red), assuming $T_{\rm s} = 300$\,K, $c_{\rm f} = 1$, $\Delta{v}_{\rm FWHM} = 30$\,km\,s$^{-1}$. The green shaded region shows the FLASH frequency band, and the grey hatched those frequencies most affected by RFI.} \label{figure:deltaX_vs_frequency}
\end{center}
\end{figure}

In \autoref{figure:deltaX_vs_frequency} we show the resulting $\Delta{X}$ as a function of observed frequency and integration time per pointing for different $N_{\rm HI}$ sensitivities, assuming a fiducial spin temperature $T_{\rm s} = 300$\,K, source covering factor $c_{\rm f} = 1$ and FWHM $\Delta{v} = 30$\,km\,s$^{-1}$. The absorption path length increases with decreasing column density sensitivity, due to fainter radio sources being included in the survey. The shape of these curves is determined by a combination of the sensitivity (SEFD) of ASKAP as a function of frequency, the underlying redshift distribution of radio sources (with a median $z \approx 1$) and the increase in comoving interval with redshift for a fixed redshift interval. The absorption path length peaks at a frequency of about 800\,MHz, where the sensitivity of ASKAP starts to reduce significantly  at lower frequencies. It is clear from this plot that the frequency range chosen for FLASH maximises the total absorption path length for the survey. Likewise, for a fixed total survey time, the discovery potential of a wide-field absorption-line survey at the sensitivity of ASKAP is optimised by maximising the sky area.

\subsubsection{Field placement and pointing centres}

\begin{figure}
\begin{center}
\includegraphics[width=\columnwidth]{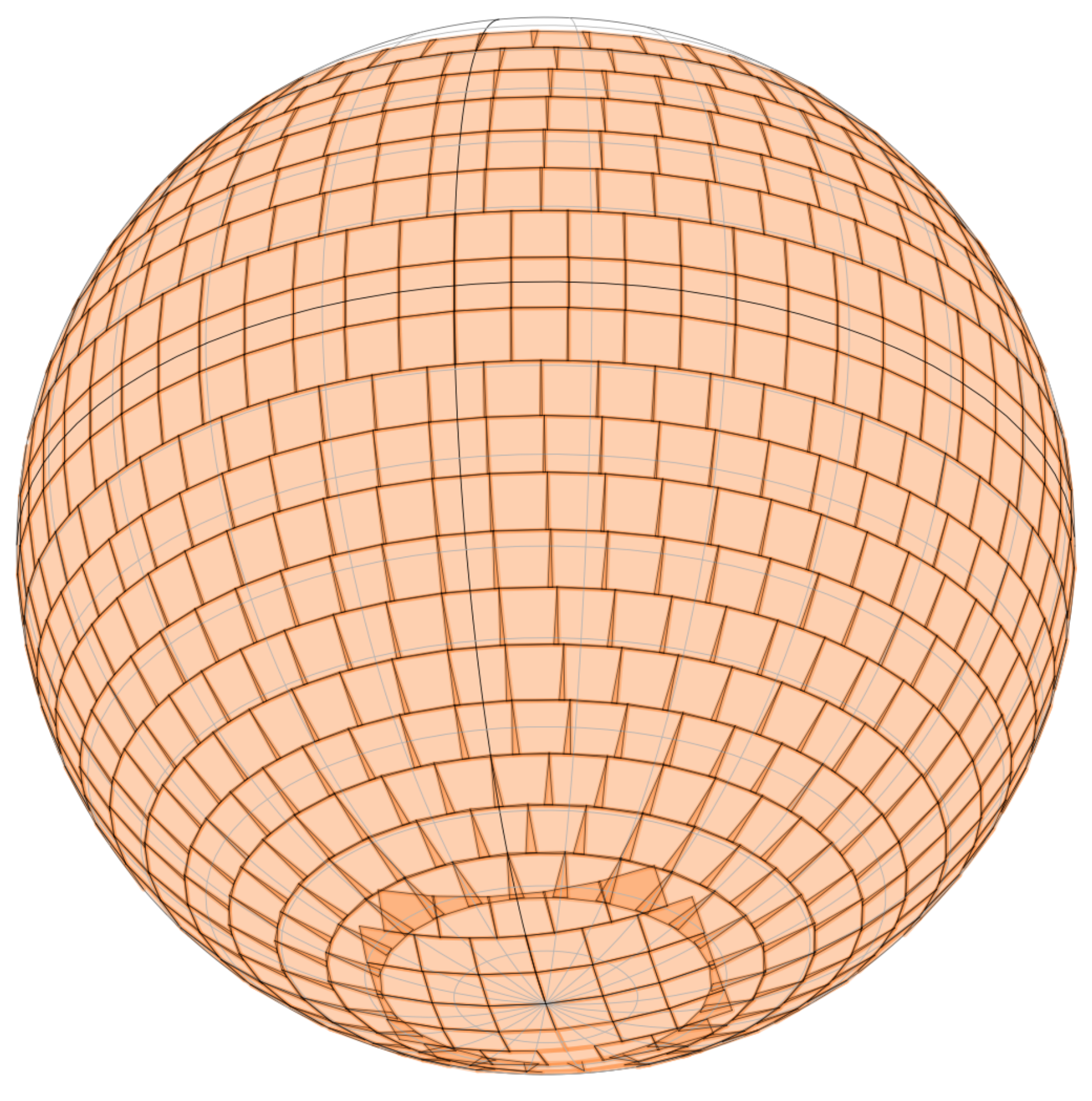}
\caption{The arrangement of the 903 planned ASKAP pointing centres for the FLASH survey, shown on the celestial sphere (adapted from Figure 3 of \citealt{McConnell:2020}). } 
\label{figure:racs_sky_coverage}
\end{center}
\end{figure}

In \autoref{figure:racs_sky_coverage} we show the positions of the pointing centres chosen for the FLASH survey. These provide an optimal sampling of the sky with uniform sensitivity. The total number of FLASH pointings is 903, equating to a total integration time for the survey of about 1806\,hr. Table \ref{table:fields} shows an indicative subset of the list of planned pointing centres for the 903 fields\footnote{The full table of pointing centres is available in the online version of the paper.} FLASH uses the same pointing centres as the 888\,MHz Rapid ASKAP Continuum Survey \citep[RACS;][]{McConnell:2020}, and the corresponding RACS field name is also shown for reference in Table \ref{table:fields}.

\begin{table*}
  \caption{Pointing centres for the fields to be used for the FLASH survey. The first ten fields are listed below, and the full list of 903 FLASH fields is available as an online table. }
  \label{table:fields}
  \begin{tabular}{rccccccccc}
    \hline
    \hline
 & & & \multicolumn{4}{c}{ --- J2000 coordinates --- } & \multicolumn{2}{c}{Galactic } \\
 No. & Field name  &  RACS field & RA & Dec & RA & Dec & $l$ &$b$ & \\
 & & & [hms] & [dms] & [deg] & [deg] & [deg] & [deg] \\
 \hline
 
1 & FLASH\_001 & RACS\_0259-85A & 02 59 49.892 & $-$85 32 47.0 & 44.958 & $-$85.547 & 300.18 & $-$30.87 &  \\
2 & FLASH\_002 & RACS\_0900-85A & 09 00 10.107 & $-$85 32 47.0 & 135.042 & $-$85.547 & 298.78 & $-$24.70 &   \\
3 & FLASH\_003 & RACS\_1459-85A & 14 59 49.892 & $-$85 32 47.0 & 224.958 & $-$85.547 & 305.51 & $-$23.33 &   \\
4 & FLASH\_004 & RACS\_2100-85A & 21 00 10.107 & $-$85 32 47.0 & 315.042 & $-$85.547 & 307.26 & $-$29.43 &   \\
5 & FLASH\_005 & RACS\_0113-80A & 01 13 38.034 & $-$80 02 45.0 & 18.408 & $-$80.046 & 301.73 & $-$37.03 &   \\
6 & FLASH\_006 & RACS\_1046-80A & 10 46 21.965 & $-$80 02 45.0 & 161.592 & $-$80.046 & 297.50 & $-$18.52 &   \\
7 & FLASH\_007 & RACS\_1313-80A & 13 13 38.034 & $-$80 02 45.0 & 198.408 & $-$80.046 & 303.93 & $-$17.22 &   \\
8 & FLASH\_008 & RACS\_2246-80A & 22 46 21.965 & $-$80 02 45.0 & 341.592 & $-$80.046 & 309.26 & $-$35.49 &   \\
9 & FLASH\_009 & RACS\_0445-80A & 04 45 52.856 & $-$80 02 37.0 & 71.470 & $-$80.044 & 292.91 & $-$31.96 &   \\
10 & FLASH\_010 & RACS\_0714-80A & 07 14 07.143 & $-$80 02 37.0 & 108.530 & $-$80.044 & 291.92 & $-$25.72 &    \\

\hline
\hline
\end{tabular}
\end{table*}


\subsection{Expected number of detections}\label{section:no_detections}

The eventual detection yield of 21-cm absorbers from FLASH will depend upon several factors, which include the distribution of \mbox{H\,{\sc i}} gas; line-of-sight spin temperatures; line widths; and the source population, including their redshifts and morphology which affects the covering factor. Significant uncertainty remains about the true distributions of these factors, and in fact determining how the 21-cm absorber population evolves with respect to the total \mbox{H\,{\sc i}} gas forms a key motivation for the survey. However, despite these uncertainties it is still instructive to estimate what we might expect to discover with FLASH. 


\subsubsection{Intervening 21-cm absorbers}

We estimate the number of detected intervening absorbers by integrating the \mbox{H\,{\sc i}} column density frequency distribution function, $f(N_{\rm HI})$, measured from previous 21-cm emission and DLA surveys over the expected comoving absorption path and column density sensitivity. Further details of this procedure are described in \autoref{section:no_intervening_absorbers}. 

In \autoref{table:nabs} we show the results for the expected total redshift path ($\Delta{z}$) and comoving absorption path length ($\Delta{X}$) for different column density sensitivities, along with the corresponding number of intervening absorbers ($\mathcal{N}_{\rm abs}^{\rm int}$) of that column density or less. In addition to our fiducial spin temperature of $300$\,K, we also give results for $T_{\rm s} = 100$ and 1000\,K covering the expected range of possible mean spin temperatures. As previously mentioned, these are highly dependent on the source covering factor $c_{\rm f}$ and line width, and so the values quoted are purely indicative. However, it is clear that FLASH is most sensitive to DLA systems that have column densities $N_{\rm HI} \sim 2 \times 10^{21}$\,cm$^{-2}$ (``super-DLAs''), and is expected to discover between several hundred and a few thousand new 21-cm absorbers. Therefore FLASH will provide a sample that is about an order of magnitude more than the current literature. In \autoref{figure:nabs_vs_frequency} we show the number of detected absorbers as a function of observed frequency, highlighting that the number of absorbers for FLASH is optimised by using the lowest frequencies available with ASKAP. 

We note that our estimate of $\Delta{z} = 6100$ for a sensitivity of $N_{\rm HI} = 2 \times 10^{20}$\,cm$^{-2}$ at $T_{\rm s} = 100$\,K is significantly less than that estimated by \cite{Gupta:2016} for FLASH (see their table 1 and figure 1). This is the result of different assumptions about the completeness and line width used in each work. In determining a detection limit for FLASH, Gupta et al. assume a relatively narrow line width of 5\,km\,s$^{-1}$ (equal to the resolution of most extragalactic 21-cm surveys), and that any feature greater than 5 times the signal-to-noise ratio will be reliably recovered with confidence. This results in a redshift path that is about 10 times that found here. Given that the distribution of known line widths is much wider than this and that we believe our estimate of the recovery of lines in data to be more accurate, our estimate is more likely to be a true representation of the final survey redshift path. 


\subsubsection{Associated 21-cm absorbers}

We predict the number of detected associated 21-cm absorbers by integrating the source redshift distribution over the redshift path that is sensitive to absorption, and then multiplying by a fixed detection rate, $\lambda_{\rm asc}$ (see \autoref{section:no_associated_absorbers} for further details of the method). As discussed in \autoref{section:current_status_observations}, the factors that determine the associated absorber detection rate are complex and require careful consideration of the properties of the sampled radio sources and their host galaxies (see \citealt{Morganti:2018} for a review). Here we adopt a fiducial rate of $\lambda_{\rm asc} = 10$\,\% for associated \mbox{H\,{\sc i}} absorption in the volume searched by FLASH. This is a factor of $2 - 3$ times lower than the typical detection rates obtained by previous targeted surveys, which typically selected samples of radio sources based on a core flux density limit (see e.g. \citealt{Maccagni:2017}). A lower detection rate is more realistic for wide-field flux-density-selected surveys. Indeed the actual detection rate could be even less than assumed here, and hence our predictions are purely indicative of the results that may be obtained from a large unbiased radio-selected survey. 

\begin{figure}[t]
\begin{center}
\includegraphics[width=\columnwidth]{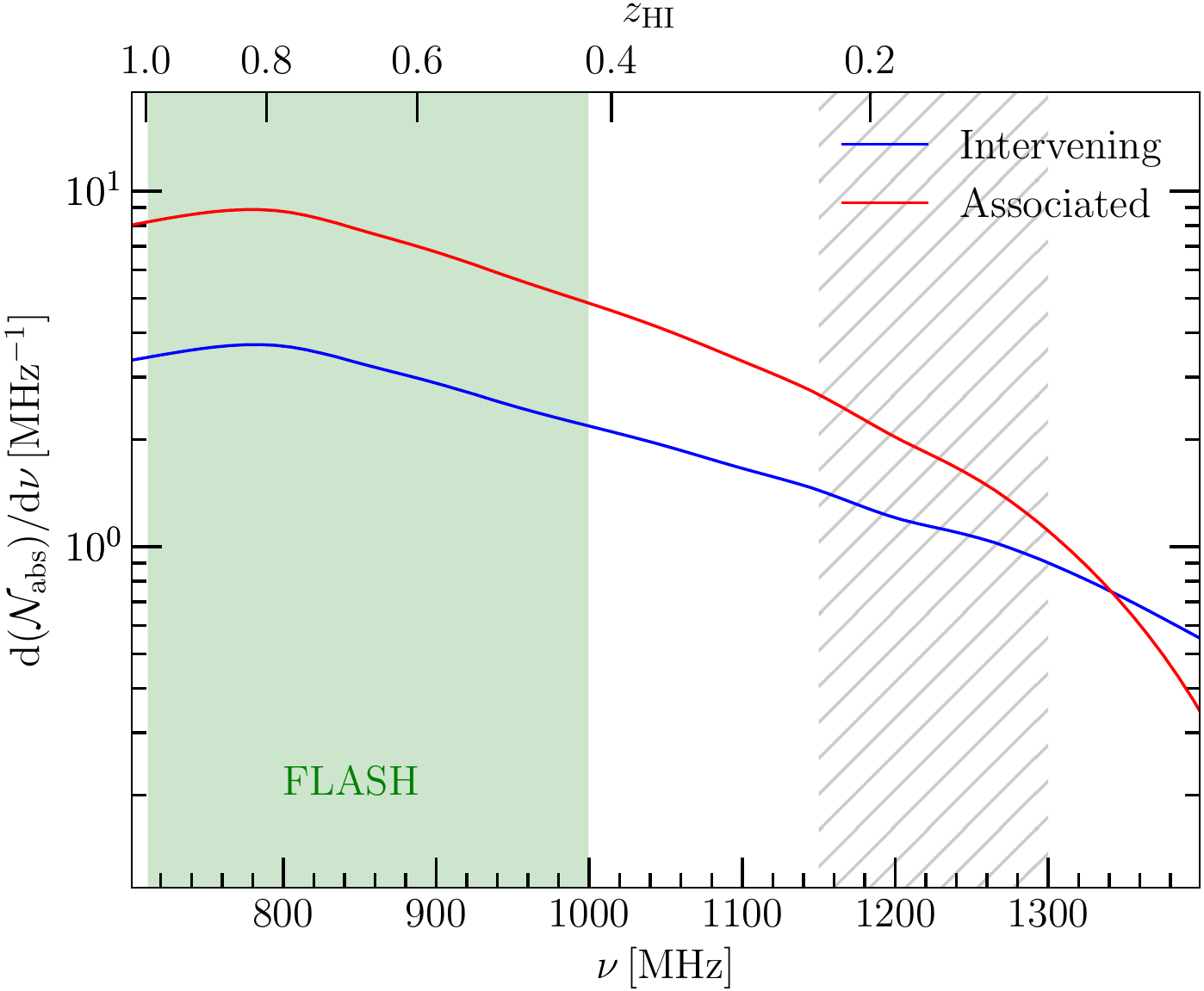}
\caption{The expected number of intervening (for $T_{\rm s} = 300$\,K, $c_{\rm f} = 1$, $\Delta{v}_{\rm FWHM} = 30$\,km\,s$^{-1}$) and associated 21-cm absorbers (for $\lambda_{\rm asc} = 10\%$, $\tau = 0.05$, $\Delta{v}_{\rm FWHM} = 120$\,km\,s$^{-1}$) detected in FLASH as a function of frequency (see text for further details). The green region shows the FLASH frequency band and the grey hatched region those frequencies most affected by RFI.} \label{figure:nabs_vs_frequency}
\end{center}
\end{figure}

We use a peak optical depth of $\tau_{\rm peak} = 0.05$ and line width $\Delta{v}_{\rm FHWM} = 120$\,km\,s$^{-1}$ for the associated absorbers (e.g. \citealt{Curran:2016a}), corresponding to a column density of $N_{\rm HI} \approx 3 \times 10^{21}$\,cm$^{-2}$ for $T_{\rm s} = 300$\,K. The sensitivity of each sight line to associated absorption is calculated assuming an unrealistic $c_{\rm f} = 1$, which is accounted for by adopting a detection rate that is significantly lower than that obtained for compact radio sources. In \autoref{figure:nabs_vs_frequency} we show the number of associated absorbers as a function of frequency, which rises considerably with \mbox{H\,{\sc i}} redshift and peaks at $z_{\rm HI} = 0.8$. This behaviour is largely governed by the source redshift distribution, except at the highest redshifts where the ASKAP sensitivity rapidly declines. This highlights that the choice of frequencies for FLASH is optimal for detecting both intervening and associated absorbers. We note that this assumes that the intrinsic detection rate is constant with redshift, which is unlikely to be the case since the cold gas content of radio galaxies will evolve with the host population (e.g. \citealt{Heckman:2014}). However, the detection of any such evolution is an important science case for the survey.

FLASH is expected to detect about 2000 associated absorbers, which is an order of magnitude greater than currently known. In their review of associated \mbox{H\,{\sc i}} absorption, \cite{Morganti:2018} predict a higher yield for FLASH of 5500 absorbers (see their table 2), based on the detection limit given by \cite{Gupta:2016} and a detection rate of $\lambda_{\rm asc} = 25\,\%$ from the results of \cite{Maccagni:2017}. Morganti et al. use the semi-empirical simulations of \cite{Wilman:2008} to create an input catalogue of radio sources, while we use the observed NVSS catalogue and apply a statistical redshift distribution. Remarkably, despite the different methods and assumptions employed, the dissagreement between these two estimates is almost entirely accounted for by the adopted detection rate. 
Finally, by taking an absorption-weighted mean over sight lines, we find that the expected flux density of sources with detected associated absorption is approximately 530\,mJy, which corresponds to a radio luminosity of $L \approx 3 \times 10^{27}$\,W\,Hz$^{-1}$ at $z = 0.8$. Therefore, we expect that most associated absorbers detected in FLASH will be associated with the most powerful radio galaxies in the Universe.

\begin{table}
\centering
\begin{threeparttable}
\caption{Estimates of the 21-cm line total redshift interval, comoving path length and number of detections for the full FLASH survey at redshifts $0.4 < z < 1$.}
\label{table:nabs}
\begin{tabular}{@{}lccc@{}}
\hline\hline 
\medskip$T_{\rm s}$ & 100\,K & 300\,K & 1000\,K \\
\hline
\multicolumn{4}{c}{\bf Intervening absorbers} \\

& \multicolumn{3}{l}{(i) Sensitivity to $\log_{10}[N_{\rm HI}/\mathrm{cm}^{-2}] = 20.3$ } \\ 
$\Delta{z}$\tnote{(a)} & 6100 & 1200 & 170 \\
$\Delta{X}$\tnote{(a)} & 12\,000 & 2200 & 320  \\ 
\medskip$\mathcal{N}_{\rm abs}^{\rm int}$\tnote{(a)} & 280 & 51 & 7 \\ 

& \multicolumn{3}{l}{(ii) Sensitivity to $\log_{10}[N_{\rm HI}/\mathrm{cm}^{-2}] = 21.3$} \\
$\Delta{z}$\tnote{(a)} & 80\,000 & 28\,000 & 6100 \\
$\Delta{X}$\tnote{(a)} & 150\,000 & 54\,000 & 12\,000 \\ 
\medskip$\mathcal{N}_{\rm abs}^{\rm int}$\tnote{(a)} & 2200 & 550 & 95 \\ 

& \multicolumn{3}{l}{(iii) All 21-cm absorption lines} \\
$\Delta{z}$\tnote{(a)} & & 230\,000 & \\
$\Delta{X}$\tnote{(a)} & & 440\,000 & \\ 
\medskip$\mathcal{N}_{\rm abs}^{\rm int}$\tnote{(a)} & 2800 & 850 & 180 \\ 
\hline
\multicolumn{4}{c}{\bf Associated absorbers} \\
\smallskip$\mathcal{N}_{\rm abs}^{\rm asc}$\tnote{(b)} & & 2000 & \\

\hline\hline
\end{tabular}
\begin{tablenotes}
\item[a] Intervening absorbers, assuming a fixed $T_{\rm s}$, $c_{\rm f} = 1$ and $\Delta{v}_{\rm FWHM} = 30$\,km\,s$^{-1}$.
\item[b] Associated absorbers, assuming $\lambda_{\rm asc} = 10$\,\%, $\tau = 0.05$ and $\Delta{v}_{\rm FWHM} = 120$\,km\,s$^{-1}$. 
\end{tablenotes}
\end{threeparttable}
\end{table}


\subsection{Comparison with other 21-cm surveys} 

\begin{table*}[t]
\caption{Comparison with other large \mbox{H\,{\sc i}} 21-cm absorption surveys. See text for details.}
\centering
\def\arraystretch{1.1}
\begin{threeparttable}
\begin{tabular}{@{}lcccccc@{}}
\hline\hline 
Survey & FLASH & MALS\tnote{(a)} & WALLABY\tnote{(b)} & SHARP\tnote{(c)} &  GBT\tnote{(d)} & ALFALFA Pilot\tnote{(e)} \\
\hline
Total sky area [$\deg^{2}$] & 34\,000 & $700 - 1000$ & 30\,940 & 2500 &  ---\tnote{($\ddagger$)} & 517 \\
$z_{\rm HI}$ & $0.4 - 1.0$ & $0 - 1.44$ &  $0 - 0.26$ & $0 - 0.26$ & $0 - 2.74$ & $0 - 0.06$ \\
$\Delta{v}_{\rm chan}$ [km\,s$^{-1}$] & $5.5 - 7.8$ & $5.5 - 13.4$ & $3.9 - 4.9$ & $7.7 - 9.7$ & $2.6 - 9.6$ & $10.4$ \\
$\sigma_{\rm chan}$ [mJy\,bm$^{-1}$] & $3.2 - 5.1$ & $0.5 - 0.6$ & $\approx 1.6$ & $\approx 1.4$ & $\approx 2 - 30$ &  $1.3 - 4.3$ \\
$\Delta{X}~(\log_{10}[N_{\rm HI}/\mathrm{cm}^{-2}] = 20.3)$ & 2200\tnote{($\ast$)} & 1600\tnote{($\ast$)} & 810\tnote{($\ast$)} & 47\tnote{($\ast$)} & 160\tnote{($\dagger$)} & 7\tnote{($\dagger$)} \\
$\Delta{X}~(\log_{10}[N_{\rm HI}/\mathrm{cm}^{-2}] = 21.3)$ & 54\,000\tnote{($\ast$)} & 17\,000\tnote{($\ast$)} & 15\,000\tnote{($\ast$)} & 990\tnote{($\ast$)} & 160\tnote{($\dagger$)} & 130\tnote{($\dagger$)} \\
\hline\hline

\end{tabular}

\begin{tablenotes}
\item[]Refs: $^{\rm(a)}$\citet{Gupta:2016}, $^{\rm(b)}$\citet{Koribalski:2020}, $^{\rm(c)}$\citet{vanCappellen:2021}, $^{\rm(d)}$\citet{Grasha:2020}, $^{\rm(e)}$\citet{Darling:2011}
\item[]$^{\ddagger}$ Targeted observations of 252 sources at various redshifts.
\item[]$^\ast$ Estimated using the method described in this work, and assuming $T_{\rm s} = 300$\,K, $c_{\rm f} = 1$, and $\Delta{v}_{\rm FWHM} = 30$\,km\,s$^{-1}$. We exclude frequencies between 1150 and 1300\,MHz due to satellite-generated RFI, which correspond to \mbox{H\,{\sc i}} redshifts between $z_{\rm HI} = 0.09$ and 0.24. 
\item[]$^\dagger$ As published, using different assumptions to the method described in this work.
\end{tablenotes}

\end{threeparttable}
\label{table:survey_comparison}
\end{table*}

In \autoref{table:survey_comparison} we compare the key survey parameters of FLASH with other planned large \mbox{H\,{\sc i}} 21-cm absorption surveys, specifically the MeerKAT Absorption Line Survey (MALS; \citealt{Gupta:2016}), the Widefield ASKAP L-band Legacy All-sky Blind surveY (WALLABY; \citealt{Koribalski:2020}) and the Search for \mbox{H\,{\sc i}} Absorption with AperTIF (SHARP; e.g. \citealt{vanCappellen:2021}, Morganti et al. in preparation). For comparison with previous work, we include the wide-field spectroscopically-blind survey by \citet{Darling:2011} using pilot data from the ALFALFA survey (\citealt{Giovanelli:2005}), and the recent targeted survey by \cite{Grasha:2020} using the Green Bank Telescope (GBT). 

These new surveys represent a significant increase in the search path for 21-cm absorbers; by covering the largest area of sky, FLASH will have the largest $\Delta{X}$ that is sensitive to high-column-density DLAs at intermediate cosmological redshifts, while MALS will target fields that contain at least one bright radio-loud quasar and will be particularly sensitive to low-column-density gas (with the caveat that $T_{\rm s}$ does not increase), probing $\mbox{H\,{\sc i}}$ across a range of environments and redshifts. 

WALLABY is an all-sky southern survey for \mbox{H\,{\sc i}} in the nearby Universe that commensally provides a low-redshift component to FLASH. Using the method described in \autoref{section:no_detections} and taking into account the RFI afflicted band between 1150 and 1300\,MHz, we expect WALLABY to add about 210 intervening and 240 associated absorbers to the FLASH catalogue (for $T_{\rm s} = 300$\,K and assuming an upper redshift $z_{\rm HI} = 0.26$). In the northern hemisphere, SHARP will also carry out a wide-field survey of absorbers in the nearby Universe. In addition to these new interferometric surveys, the Five-hundred-metre Aperture Spherical Telescope (FAST) will carry out drift scan surveys that are expected to detect at least several hundred 21-cm absorbers in galaxies out to $z \approx 0.35$ (\citealt{Zhang:2021}).


\section{Science Goals}\label{section:science_goals}

FLASH is designed to achieve several key science goals based on the detection and characterisation of the cold neutral gas and radio continuum in galaxies at cosmological distances. In this section we discuss these science goals in more detail.


\subsection{The nature of \mbox{H\,{\sc i}} absorption-selected galaxies}\label{section:nature_absorption-selected_galaxies}

We expect to be able to identify the host galaxies associated with many of the \mbox{H\,{\sc i}} absorption lines detected in the FLASH survey - either by using existing optical and infrared surveys\footnote{The WISE mid-IR survey \citep{Cutri:2014} covers the whole FLASH survey area, while optical photometry from SDSS \citep{Blanton:2017} is available for much of the region north of dec 0$^o$ and the forthcoming LSST survey \citep{Ivezic:2019} will provide deep optical photometry for most of the sky south of dec 0$^o$.} or through a program of follow-up imaging and spectroscopy with optical telescopes \citep[see e.g.][]{Sadler:2020}. 

In particular, FLASH can provide us with a representative sample of genuinely `\mbox{H\,{\sc i}}-selected' galaxies at $0.4<z<1.0$. Such a sample will show how \mbox{H\,{\sc i}} is distributed in and around galaxies in this redshift range, and allow us to relate \mbox{H\,{\sc i}} to star formation at the same epoch. The FLASH absorption results can also help to guide the selection of galaxy samples for \mbox{H\,{\sc i}} emission-line stacking experiments (see section \ref{section:stacking}). 

\begin{figure*}[t]
\begin{center}
\includegraphics[width=0.99\columnwidth]{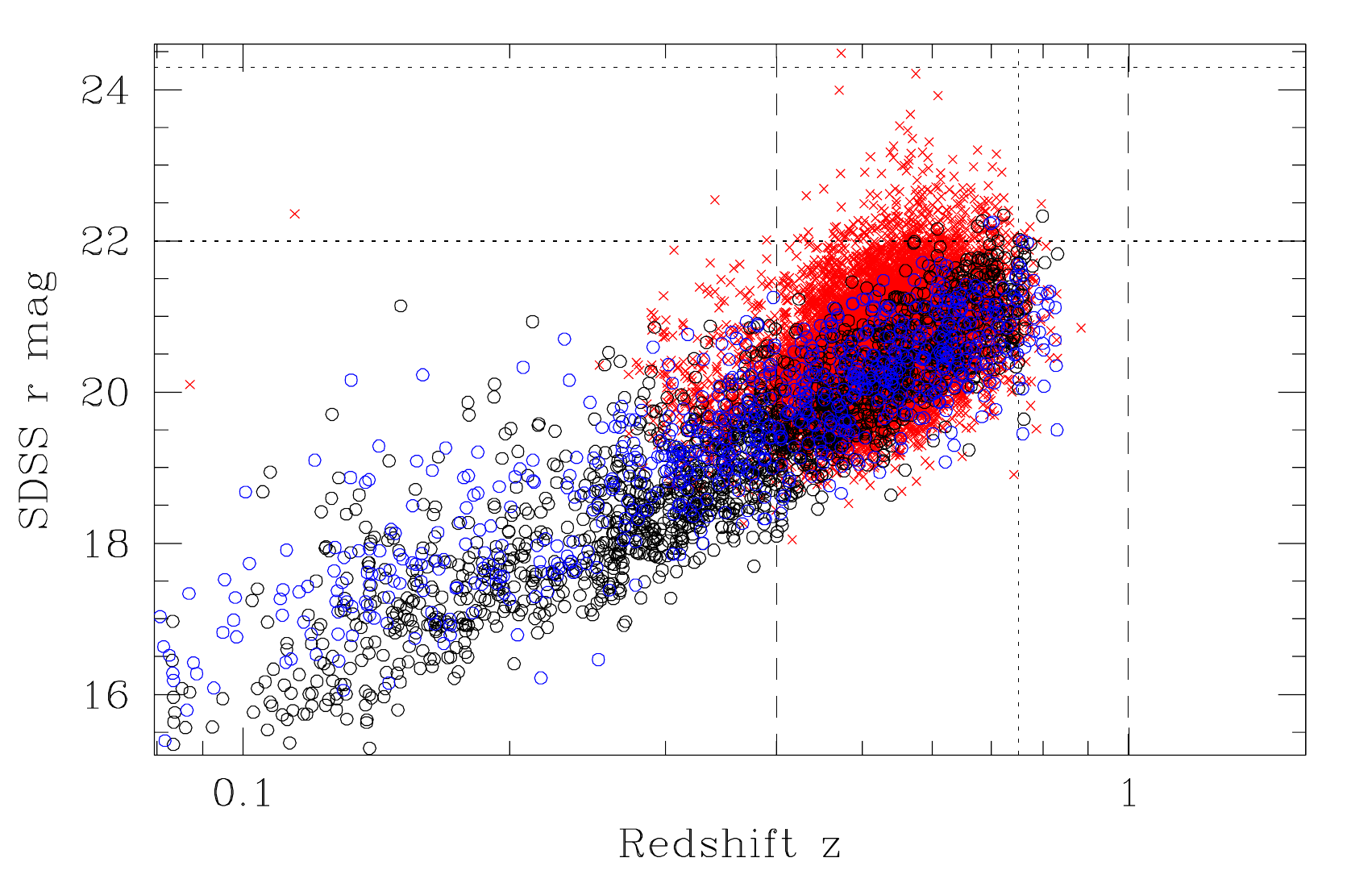} 
\includegraphics[width=1.0\columnwidth]{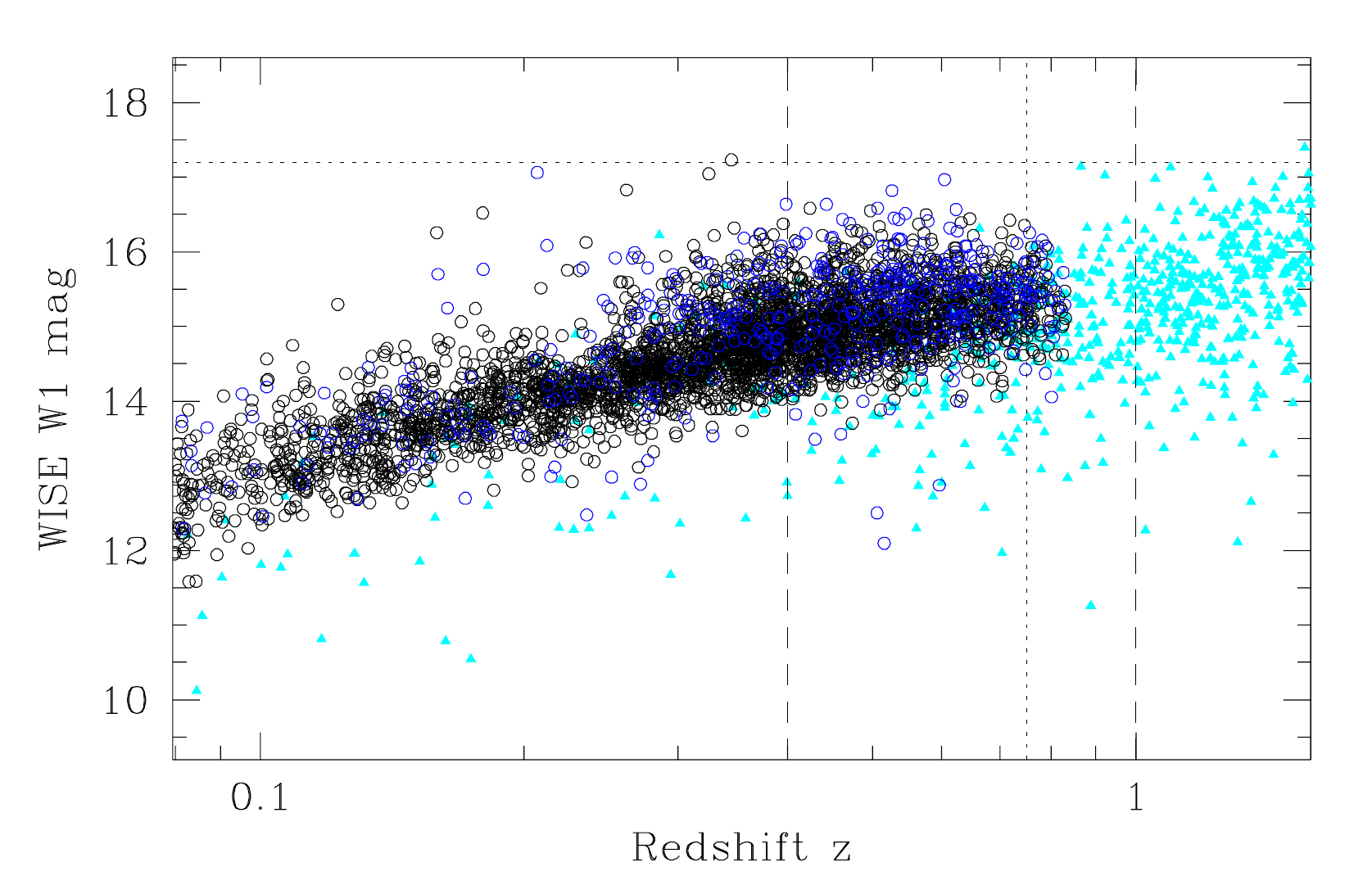} 
\caption{
Optical and mid-IR properties of the hosts of radio AGN similar to those that will be observed in FLASH. In both plots, vertical dashed lines show the redshift range covered by the main FLASH survey ($0.4 < z < 1$) and the vertical dotted line shows the point at which current large-area spectroscopic galaxy surveys start to become incomplete ($z \sim 0.75$).
{\it Left:} Observed SDSS r-band magnitude versus redshift for several galaxy classes. Red points show luminous red galaxies \citep[LRGs, spectroscopic redshifts from the 2SLAQ Survey,][]{Cannon:2006}, while black and blue points show low-excitation and high-excitation radio galaxies, respectively \citep[LERGs/HERGs, spectroscopic redshifts from][]{Ching:2017}. The horizontal dotted lines show the photometric limit of the SDSS catalogue (r=22), and the expected single-visit depth for LSST (r=24.3). 
{\it Right:} WISE W1 band (3.4\,$\mu$m) magnitude versus redshift for low-excitation radio galaxies (LERGs, black circles),  high-excitation radio galaxies (HERGs, blue circles) and radio-loud QSOs (cyan triangles), all from the  \cite{Ching:2017} catalogue. The horizontal dotted line shows the completeness limit of the WISE catalogue (W1=17.2). 
 } \label{figure:flash_optical}
\end{center}
\end{figure*}


\subsubsection{Radio-loud AGN with associated 21-cm absorption}

No optical pre-selection is used in the FLASH survey, and we can therefore detect \mbox{H\,{\sc i}} gas even when no optical galaxy is visible. However, we expect most {\it associated}\ 21-cm absorbers to lie in the kinds of massive galaxies that host radio-loud AGN, i.e. close to the locus of the black points in Figure \ref{figure:flash_optical}. QSOs with associated \mbox{H\,{\sc i}} absorption may be even brighter than this, while some host galaxies of high-excitation radio sources (HERGs) could be one or two magnitudes fainter \citep[e.g.][see the blue points in Figure \ref{figure:flash_optical}]{Ching:2017}. Thus most associated absorbers detected by FLASH should have an optical/IR counterpart visible in WISE mid-IR images, and coincident with the radio position (or the radio centroid for an extended source). 

\begin{figure}
    \centering
    \includegraphics[width=\columnwidth]{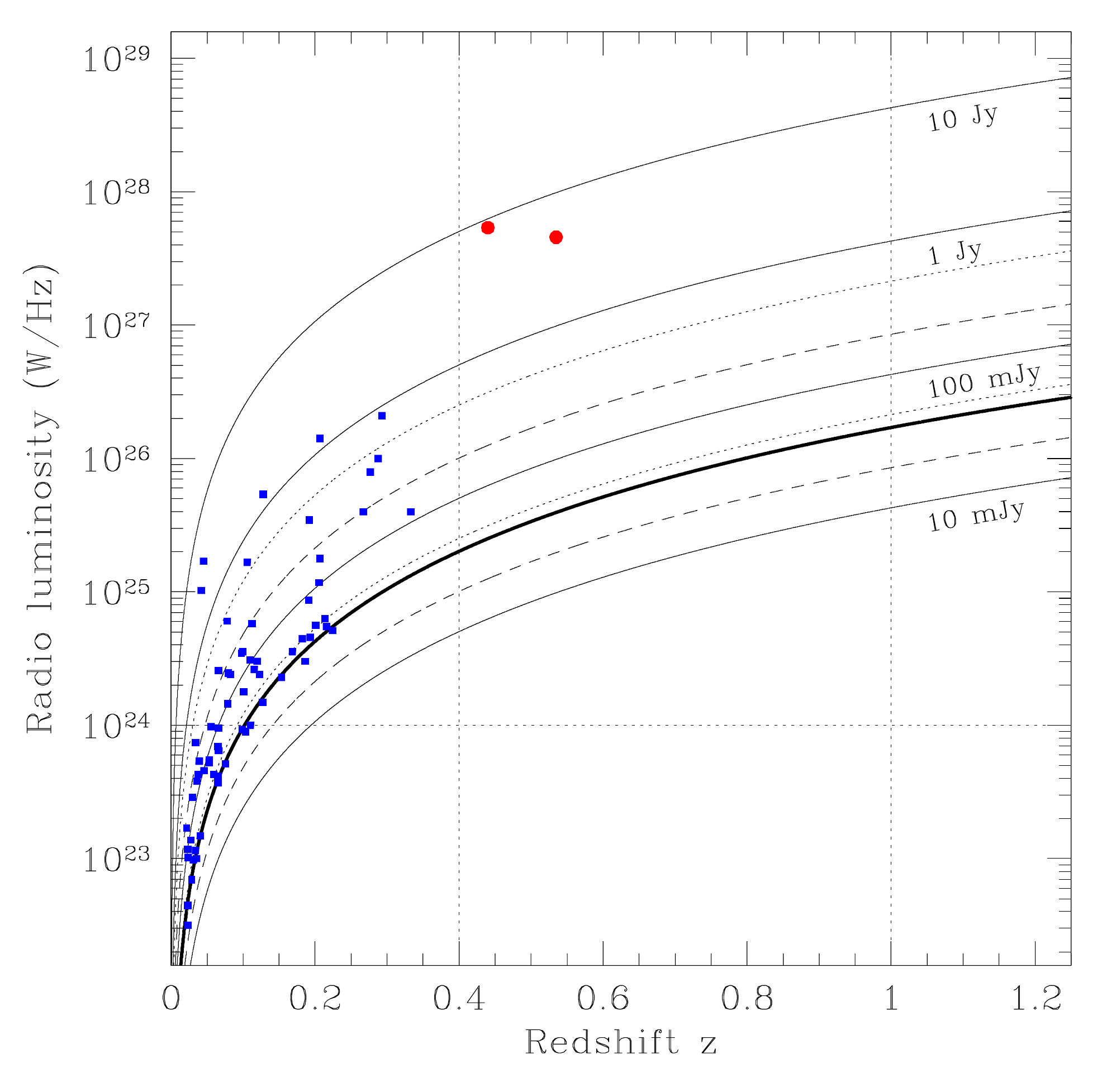}
    \caption{Radio luminosities of some representative objects in which associated \mbox{H\,{\sc i}} has been detected. Blue points show detections from the lower-frequency samples published by \cite{Maccagni:2017} and \cite{Murthy:2021}, while red points show two detections from ASKAP commissioning data (PKS\,1740-517, \cite{Allison:2015} and PKS\,1829-718, 
    \cite{Glowacki:2019}). The thick line at a flux density of 40\,mJy indicates an approximate detection limit for absorption systems in the FLASH survey.  }
    \label{fig:zlum}
\end{figure}

Figure \ref{fig:zlum}\ compares the radio luminosity of the radio AGN in which \mbox{H\,{\sc i}} was detected by \cite{Maccagni:2017} and \cite{Murthy:2021} at $z < 0.4$\ with two associated absorption lines detected in ASKAP commissioning at $z \sim 0.5$ \citep{Allison:2015, Glowacki:2019}. These early ASKAP detections are associated with extremely bright radio sources with radio luminosities above 10$^{27}$\,W\,Hz$^{-1}$, but FLASH should also be able to detect \mbox{H\,{\sc i}} absorption (and especially lines with high optical depth) in sources as faint as 40\,mJy. Thus FLASH can probe some sources with radio luminosities between 10$^{25}$ and 10$^{26}$\,W\,Hz$^{-1}$ in addition to the population of powerful sources above 10$^{26}$\,W\,Hz$^{-1}$ -- which are relatively rare in the local Universe but far more common at $z > 0.4$ \citep{Pracy:2016}. 


\subsubsection{Host galaxies of intervening 21-cm absorption}

The host galaxies of {\it intervening}\ 21-cm absorbers could in principle be associated with galaxies of almost any magnitude. The optical luminosity function of galaxies is very broad \citep[e.g.][]{Blanton:2001}, with absolute magnitudes ranging from M$_r = -23$ for the most massive and luminous galaxies to M$_r = -16$ or even fainter for dwarf galaxies. The `knee' of the galaxy luminosity function in the local Universe is at M$_r = -20.8$, but the observed r-band magnitude at higher redshift will depend on both M$_r$ and the k-correction. 

Since the radius of the \mbox{H\,{\sc i}} disk in nearby galaxies is known to scale extremely well with \mbox{H\,{\sc i}} mass, however \citep{Wang:2016} we would expect a sample of \mbox{H\,{\sc i}}-selected galaxies to trace the \mbox{H\,{\sc i}} mass function reasonably well. Since \mbox{H\,{\sc i}} mass roughly scales with galaxy luminosity \citep{Maddox:2015}, we might expect most galaxies associated with intervening \mbox{H\,{\sc i}} absorbers to be late-type galaxies with stellar mass above $\sim10^9$\,M$_\odot$ \citep[e.g.][]{Rodriguez:2020}. 
Such galaxies should be detectable in targeted follow-up optical imaging if they are not already visible in archival imaging surveys. 

The impact parameter for intervening 21-cm absorbers is generally expected to be less than 20 kpc \citep[e.g.][]{Borthakur:2016, Curran:2016b, Dutta:2017a, Curran:2020} even though the \mbox{H\,{\sc i}} disks of gas-rich galaxies may extend out as far as 60\,kpc at lower \mbox{H\,{\sc i}} column density ($N_{\rm HI} \sim 10^{18}$\,cm$^{-2}$, \citealt{Bland-Hawthorn:2017}). For impact parameters as high as 10 -- 20\,kpc, the intervening galaxy may be offset by several arcsec from the radio position. If the background radio source is an optically bright QSO and the impact parameter is low, this may also complicate the identification of the intervening galaxy. Despite these challenges, however, identifying the host galaxies of most FLASH detections should be tractable with a suitably-designed follow-up program. 

\subsubsection{Distinguishing associated and intervening absorbers}

\autoref{table:distinguishing_absorbers} summarises several ways of  distinguishing associated and intervening \mbox{H\,{\sc i}} absorbers, depending on the additional information is available. 

The distinction is fairly straightforward when an optical identification and spectroscopic redshift are available for the radio source against which the absorption line is detected. 

When the redshift of the radio source is unknown, intervening absorbers may still be identified if there is a significant astrometric offset between the radio source position and the nucleus of the intervening galaxy. For example, \cite{Allison:2020} used data from the GAMA survey to identify an ASKAP \mbox{H\,{\sc i}} line detection as an intervening absorber in the outer regions (impact parameter 17\,kpc) of a massive early-type galaxy. In this case the radio-optical offset was 2.5\,arcsec - significantly higher than the combined uncertainties in the optical and radio positions. 

Finally, machine learning techniques \citep{Curran:2016a, Curran:2021} have the potential to distinguish intervening and associated absorption lines based on radio spectral-line data alone. Current classifiers have a success rate of $\sim80$\% \citep{Curran:2021}, and it is possible that the accuracy can be improved in future when a larger training set of absorbers becomes available.  

\begin{table*}[t]
    \centering
    \caption{Methods for distinguishing associated and intervening \mbox{H\,{\sc i}} absorption absorbers}
    \begin{tabular}{llll}
\hline
\hline
Parameter & Associated \mbox{H\,{\sc i}} & Intervening \mbox{H\,{\sc i}} & Notes \\
          & absorption    & absorption & \\
\hline
Redshift  & $z_{\rm HI} \approx z_{\rm opt}$ & $z_{\rm HI} < z_{\rm opt}$ & Requires optical ID and redshift $z_{\rm opt}$\\
   &  & & for background radio source \\
&& \\
Astrometry & Optical position & Optical position & Requires optical ID for absorber host and \\
           & matches radio centroid & offset from radio  & sub-arcsec position for background radio source \\
&& \\
\mbox{H\,{\sc i}} profile & \multicolumn{2}{l}{Based on machine learning algorithms } & Requires good S/N in \mbox{H\,{\sc i}} line profile \\
           & \citep{Curran:2021} & & \\ 
\hline
\hline
    \end{tabular}
    \label{table:distinguishing_absorbers}
\end{table*}


\subsection{Cosmological evolution of the $\mbox{H\,{\sc i}}$ gas in galaxies} 

\begin{figure*}
\begin{center}
\includegraphics[width=\textwidth]{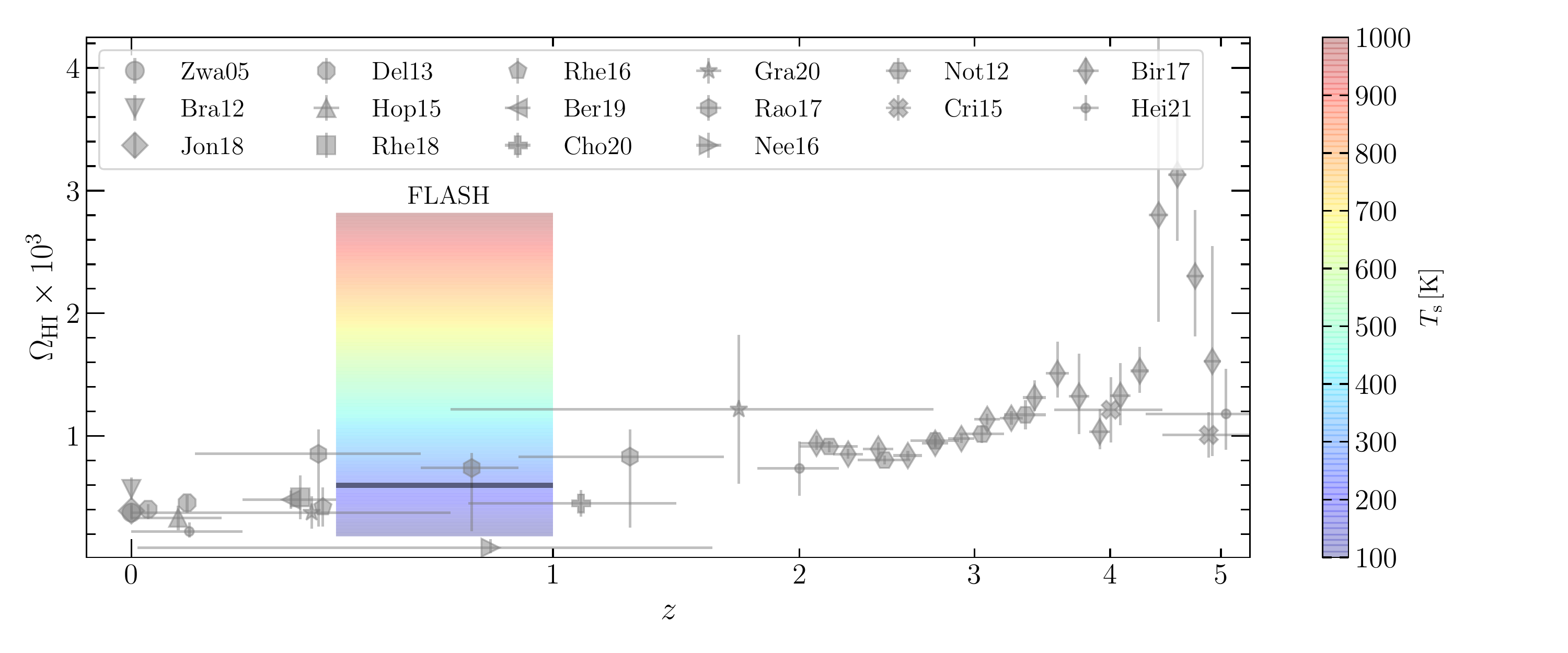}
\caption{Various measurements of the cosmological mass density in \mbox{H\,{\sc i}} gas as a function of redshift. The horizontal black bar represents the expected FLASH measurement from intervening 21-cm absorbers, assuming $T_{\rm s} = 300$\,K and $c_{\rm f} = 1$. The vertical extent of this black bar represents the standard deviation due to sample variance. The coloured region indicates how this measurement depends on the assumed harmonic mean $T_{\rm s}$ of the absorber population, which can then be inferred by comparison with other measurements (see text for details). The selected literature measurements include \emph{21-cm emission}: Zwa05 -- \citet{Zwaan:2005b}, Bra12 -- \citet{Braun:2012}, Jon18 -- \citet{Jones:2018}; \emph{21-cm stacking}: Del13 -- \citet{Delhaize:2013}, Hop15 -- \citet{Hoppmann:2015}, Rhe18 -- \citet{Rhee:2018}, Rhe16 -- \citet{Rhee:2016}, Ber19 -- \citet{Bera:2019}, Cho20 -- \citet{Chowdhury:2020a}; \emph{21-cm absorption}: Gra20 -- \citet{Grasha:2020}, assuming $T_{\rm s}/c_{\rm f} = 175$\,K; \emph{Damped Lyman-$\alpha$ Absorbers}: Rao17 -- \citet{Rao:2017}, Nee16 -- \citet{Neeleman:2016}, Not12 -- \citet{Noterdaeme:2012}, Cri15 -- \citet{Crighton:2015}, Bir17 -- \citet{Bird:2017}; and \emph{[CII] 158-$\mu$m emission}: Hei21 -- \citet{Heintz:2021}. All measurements have been corrected to a common definition, with no helium contribution, and the DLA measurements have been corrected by a further factor of 1.2 to account for sub-DLA gas (\citealt{Berg:2019}).} \label{figure:omegaHI_vs_z}
\end{center}
\end{figure*}

The cosmic star formation rate (SFR) density evolves strongly with redshift, rapidly accelerating in the early Universe, peaking at $z \approx 1.5 - 2.5$, and then declining by a factor 10 -- 15 to the present epoch (e.g. \citealt{Hopkins:2006, Madau:2014, Driver:2018}). Understanding the processes that drive this evolution over cosmic time is a major endeavour of modern astrophysics. The availability of cold neutral gas in galaxies is clearly important for the formation of self-gravitating clouds of dense molecular gas within which star formation can occur (\citealt{McKee:2007}). 

Observations of the bulk tracers of molecular gas - CO emission (e.g. \citealt{Decarli:2019, Decarli:2020, Lenkic:2020, Fletcher:2021, Riechers:2020a, Riechers:2020b}), supplemented by far-infrared and mm-wavelength observations of the dust continuum (e.g. \citealt{Berta:2013, Scoville:2017, Magnelli:2020}) -- support evolutionary models that mirror that of the SFR density. By contrast, the mass density in atomic hydrogen ($\Omega_{\rm HI}$), which traces the extended reservoir of neutral gas in galaxies, has only declined by at most a factor $\sim 2$ since the SFR peak (see \autoref{figure:omegaHI_vs_z}). This suggests that as the diffuse atomic gas is converted to denser clouds of molecular gas, it is replenished by accretion from the intra/circumgalactic medium. The decline in the SFR and molecular gas densities is therefore indicative of a decline in the gas accretion rate of galaxies (e.g. \citealt{Peroux:2020, Tacconi:2020, Walter:2020}).   

Although we have a reasonable idea of the global behaviour of \mbox{H\,{\sc i}} over the cosmic history, important details are missing. Measurements of $\Omega_{\rm HI}$ at intermediate cosmological redshifts ($0.1 < z < 2$) are not as complete as in the nearby or more distant Universe. These are obtained primarily from DLA samples observed at UV-wavelengths (\citealt{Neeleman:2016, Rao:2017}) and statistical detections of 21-cm emission by stacking at the positions of known optically-bright galaxies (\citealt{Delhaize:2013, Hoppmann:2015, Rhee:2016, Rhee:2018, Bera:2019, Chowdhury:2020a}). The UV-DLA results are limited by the sample size and possible systematic errors associated with \mbox{Mg\,{\sc ii}}-selection. The 21-cm stacking results have better measurement uncertainties, and seem to agree within the uncertainties, but they are also limited to the sample by which they are selected. 


\subsubsection{Cold gas evolution from 21-cm absorbers}

21-cm absorption-line surveys provide an important complementary measurement of the \mbox{H\,{\sc i}} content of galaxies at these redshifts. If the background sources are selected based purely on their radio properties then they are free of any potential optical selection bias. This technique was recently demonstrated by \cite{Grasha:2020}, who carried out a survey for 21-cm absorption towards 252 compact radio sources with the GBT telescope, covering redshifts between $z = 0$ and 2.4. They successfully detected 10 absorbers and, assuming $T_{\rm s}/c_{\rm f}= 175$\,K, were able to measure $\Omega_{\rm HI}$ to a comparable precision with other methods at the same redshifts (see further discussion on the spin temperature below). 

In \autoref{figure:omegaHI_vs_z} we show the measurement of $\Omega_{\rm HI}$ expected from FLASH, which is based on the expected number of intervening 21-cm absorbers shown in \autoref{table:nabs} (see \autoref{section:omega_HI} for a description of the method). Given the number of intervening 21-cm absorbers expected to be detected, the standard deviation due to sample variance will only be only a few\,per\,cent over the redshift interval of the survey (as indicated by the width of the bar in \autoref{figure:omegaHI_vs_z}), which is a marked improvement on previous measurements at these redshifts. However, this measurement is strongly dependent on the assumed value of the spin temperature and covering factor for each absorber. Therefore, we also show how it varies with the typical range of spin of temperatures seen in DLAs, indicating how such a measurement can be used to infer the gas temperature (see below for further discussion).

The unknown source covering factor can be overcome by selecting only background sources with known compact morphologies that are smaller than the expected angular scale of the foreground absorber ($\Delta{\theta} \lesssim 10$\,mas at $z \sim 1$; \citealt{Braun:2012}). If enough spatial information is known about the individual objects in the sample, then a proxy for $c_{\rm f}$ can be determined by taking the ratio of the total to compact radio flux density at frequencies close to that expected for the redshifted 21-cm line (e.g. \citealt{Kanekar:2009b}). In the case of FLASH, the number of radio source targets is likely to be $\sim 100\,000$, rendering such a targeted high-resolution campaign difficult. Alternatively, one can assume a covering factor probability distribution for a given sample of background sources, which is then used as a prior for any future inference about the optical depth of the absorber (e.g. \citealt{Allison:2016b, Allison:2021}). 

Less is known about the spin temperature in individual absorbers. For a fixed column density and source covering factor, the equivalent width of the 21-cm absorption line is inversely related to the \mbox{H\,{\sc i}} spin temperature (see \autoref{equation:column_density}). Hence detections of 21-cm absorption are weighted to line-of-sight gas that contains a greater fraction of the denser CNM (\citealt{Wolfire:2003}), where molecular cloud and star formation occur (\citealt{Krumholz:2009a}). Crucially, this means that 21-cm absorption surveys provide an important probe of the colder neutral gas in galaxies.

Direct measurement of the spin temperature in 21-cm absorbers can be achieved if the \mbox{H\,{\sc i}} column density is known by other means. The inferred spin temperature is then an $N_{\rm HI}$-weighted harmonic mean over the line-of-sight components of the neutral gas. In the nearby Universe this can be achieved by simultaneously detecting 21-cm emission and absorption in a foreground galaxy, with a few examples in the literature (e.g. \citealt{Reeves:2016, Borthakur:2016, Gupta:2018}). However, future wide-field surveys (e.g. WALLABY; \citealt{Koribalski:2020} and SHARP; \citealt{vanCappellen:2021}, Morganti et al. in preparation) are expected to increase this sample by a few orders of magnitude. 

At cosmological distances the spin temperature can instead be obtained by simultaneously detecting 21-cm and Lyman-$\alpha$ absorption, requiring a sample of radio-loud UV or optically-selected quasars. \cite{Kanekar:2014a} compiled the literature sample into a single study, finding that the spin temperatures of DLAs at $z > 2.4$ are higher than those at lower redshifts (at 4-$\sigma$ significance). Kanekar et al. also obtained an anti-correlation between the spin temperature and gas-phase metallicity of DLAs (at 3.5-$\sigma$ significance), suggesting that in the early Universe galaxies were depleted of the metals required to form CNM via fine structure cooling. 

Such a direct study of the spin temperatures in FLASH-detected 21-cm absorbers would be very challenging since the radio source would need to be sufficiently bright at UV-wavelengths to detect Lyman-$\alpha$ absorption using the \emph{Hubble Space Telescope}. However, we can instead use the statistical power of such a large sample by comparing the number of 21-cm absorbers detected in the survey with that expected from the $N_{\rm HI}$ distribution function measured from 21-cm emission and DLA surveys. This can then be used to obtain a statistical measurement of the $N_{\rm HI}$-weighted harmonic mean spin temperature in galaxies at cosmological redshifts (\citealt{Darling:2011, Allison:2016b, Grasha:2020, Allison:2021}), thereby enabling strong constraints to be placed on the evolution of the physical state of cold gas in galaxies over the past 8 billion years.


\subsubsection{Statistical detection of HI emission from star-forming galaxies}\label{section:stacking}

The weak flux of \mbox{H\,\sc i} emission from galaxies makes it difficult to detect at anything but low redshifts with current telescopes ($z \lesssim 0.3$). To extend the measurements to higher redshift the signal from multiple galaxies has been stacked in radio observations using the known optical position and redshift of galaxies. This decreases the noise in the measurement by the square root of the number of coadded galaxies. Notably, \mbox{H\,\sc i} stacking has been used by \citet{Lah:2007}, \citet{Lah:2009}, \citet{Rhee:2013}, \citet{Rhee:2016}, \citet{Rhee:2018}, \citet{Bera:2019} and \citet{Chowdhury:2020a}. The most recent result is by \citet{Chowdhury:2021} where they measured the average \mbox{H\,\sc i} emission signal from 2841 blue star-forming galaxies at $z = 1.18 - 1.39$ using 400~hours using the Giant Metrewave Radio Telescope. FLASH only has observations of 2 hours per field but it observes an extremely large area covering a large redshift range. This means that there are a large number of galaxies that can be used in the stacking analysis, compensating for the smaller integration time per field.  

\begin{figure}
\begin{center}
\includegraphics[width=\columnwidth]{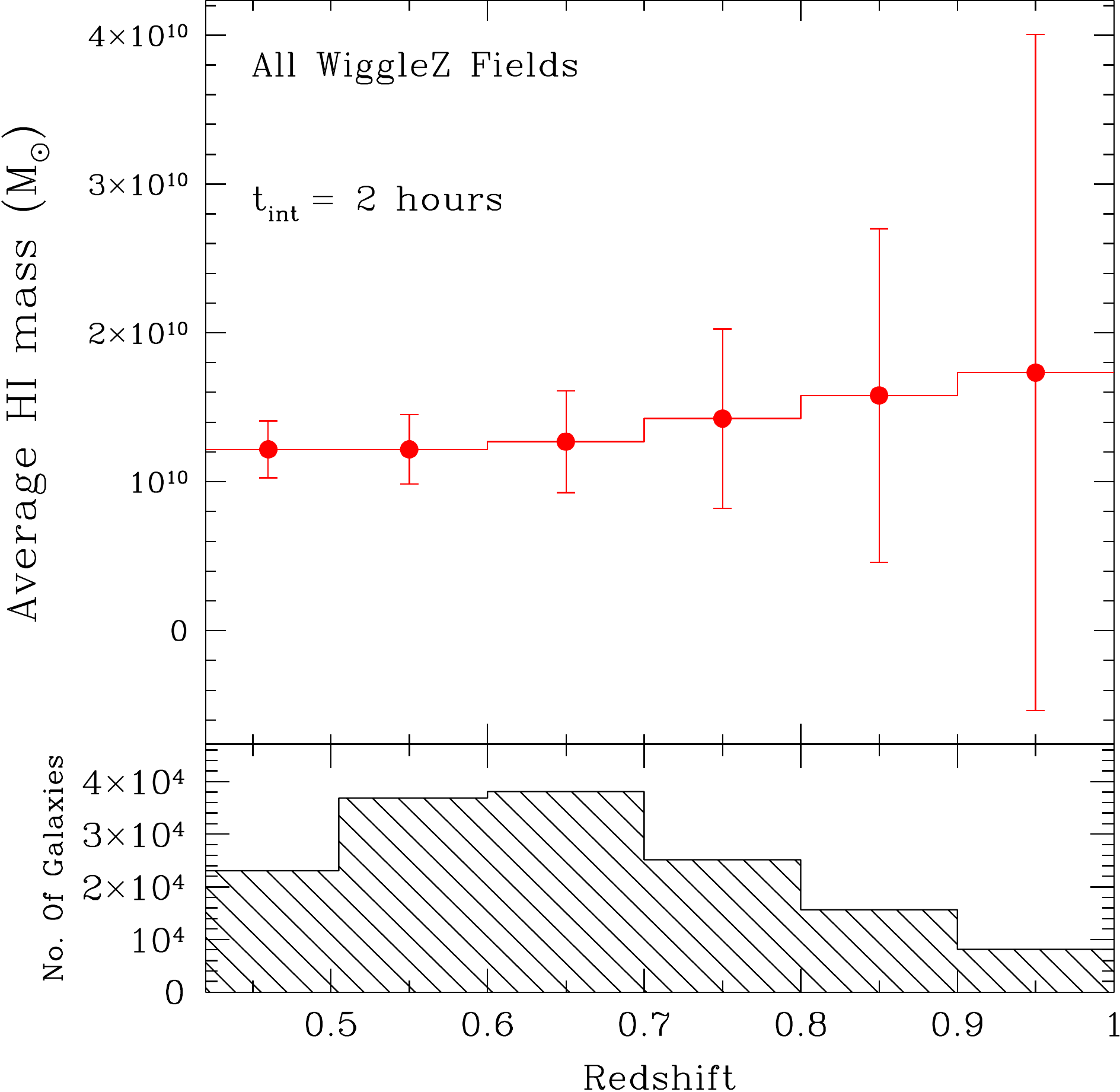}
\caption{The estimated \mbox{H\,\sc i} signal from coadding the WiggleZ in the FLASH data versus redshift.  The bottom of the plot shows the number of WiggleZ galaxies coadded in each redshift bin.  The top of the plot shows the estimated  average \mbox{H\,\sc i} for these galaxies along with the expected error in the measured signal based off the parameters of FLASH.}
\label{figure:stacking_wigglez}
\end{center}
\end{figure}

WiggleZ is an optical spectroscopic survey of star forming emission line galaxies carried out at the Anglo-Australian Telescope at Siding Spring, Australia between August 2006 and January 2011 \citep{Drinkwater:2010}. It covers 1000~deg$^2$ with redshifts in the range $z < 1$ and median $z = 0.6$. The FLASH survey covers the whole area on the sky and there are 152,057 redshifts that lie within the FLASH \mbox{H\,\sc i} redshift range. An estimate of how well stacking WiggleZ galaxies in the FLASH data is shown in \autoref{figure:stacking_wigglez}. The estimated average \mbox{H\,\sc i} mass is determined by assuming that the WiggleZ galaxies in each redshift bin are the largest \mbox{H\,\sc i} mass galaxies in the volume probed, and then using the $z = 0$ \mbox{H\,\sc i} mass function of \citet{Zwaan:2005} to allocate an \mbox{H\,\sc i} mass. A correction for the completeness of the WiggleZ survey is included. This does not consider any evolution in the \mbox{H\,\sc i} mass function which would increase the \mbox{H\,\sc i} mass measured. The stacked observations will probe the high mass end of the \mbox{H\,\sc i} mass function. Other \mbox{H\,\sc i} surveys with their smaller area coverage will only have a few of these large galaxies. The uncertainties on the average \mbox{H\,\sc i} mass are based on the parameters of FLASH and the number of galaxies stacked. The uncertainties at the low redshift end are reasonable but they rapidly become large at the high redshift end due to fewer redshifts, weaker \mbox{H\,\sc i} 21-cm flux from the galaxies, and lower ASKAP sensitivity at the relevant frequencies.  It should be noted that \citet{Li:2021} made a \mbox{H\,\sc i} intensity mapping measurement at $0.73 < z < 0.78$ using WiggleZ galaxies with the Parkes telescope. The limited redshift range was due to RFI at this site. The redshift range probed by ASKAP that is reasonably constraining on the \mbox{H\,\sc i} mass is a regime that has significant RFI at other telescope sites, thereby making the measurement relevant. There should be sufficient signal to noise at these lower redshifts to divide the sample into subsamples and measure their average \mbox{H\,\sc i} mass. In particular, determining how the average \mbox{H\,\sc i} mass varies with star formation rate at these redshifts will be of significant interest.

Another optical redshift survey that significantly overlaps with the FLASH area coverage is DESI (Dark Energy Spectroscopic Instrument) Survey  \citep{DESI_Collaboration:2016}.  This survey is being carried out with the 4-meter Mayall Telescope at Kitt Peak National Observatory. The survey will cover $14\,000\,\deg^{2}$ spanning from the equator up to some fields north of $\delta = +40\,\deg$. The overlap in sky area with FLASH is therefore large. Their ELG (Emission Line Galaxy) sample covers redshifts $z = 0.6$ to $1.6$ with 1220 objects per $\deg^2$ for a total of 17.1 million targets. This survey has started taking data and has a 5-year lifetime overlapping well with the FLASH observation schedule. The stacked \mbox{H\,\sc i} signal using DESI galaxies should be significantly better than that made with WiggleZ galaxies.

It should be noted that stacking can also be carried out on the radio continuum of galaxies, providing a measure of the star formation rate without dust contamination. In this case, AGN contamination would need to be taken into account, but this should be a solvable problem.  


\subsection{Co-evolution of powerful radio-loud AGN and their hosts}\label{section:coevolution_agn}

\subsubsection{Feedback and triggering}

It is well-established that the properties of central supermassive black holes are tightly linked to the properties of their host galaxies (e.g. \citealt{Kormendy:1995, Magorrian:1998, Ferrarese:2000}), with the connection generally attributed to AGN feedback processes \citep{DiMatteo:2005,Croton:2006,Weinberger:2017}. However, our understanding of the physical mechanisms responsible for AGN fuelling and feedback is still limited by the lack of observational evidence, particularly on scales close to the central AGN. In cases where the \mbox{H\,{\sc i}} 21-cm absorption line is at the same redshift as the radio continuum source (i.e. associated 21-cm absorbers, see Section \ref{section:observations_associated_absorbers}), this can provide insight into the interaction between the neutral gas and central radio AGN. \mbox{H\,{\sc i}} absorption studies can provide a unique perspective as they directly trace the gas at the position of the AGN, are more sensitive to the cooler gas and have the potential to probe gas at higher spatial resolution.   

The kinematics of the \mbox{H\,{\sc i}} absorption line can separate whether the gas is infalling \citep{Maccagni:2014, Tremblay:2016}, pointing towards a gas reservoir capable of fueling the central nucleus, or outflowing \citep{Morganti:1998, Gereb:2015, Morganti:2013a}, providing observational evidence of AGN-driven feedback. Low-redshift H{\sc i} absorption studies suggest that fast jet-driven outflows of H{\sc i} gas (with velocities over 1000\,km\,s$^{-1}$) dominate the feedback processes in powerful radio galaxies \citep{Morganti:2005, Mahony:2013} with studies of the multi-phase gas properties showing that the bulk of the mass in outflow is often in the neutral or molecular form \citep{Tadhunter:2014, Mahony:2016, Morganti:2018}. While this has been well studied in the nearby Universe, it is crucial to extend these studies to higher redshifts when feedback processes have the most impact \citep{Hopkins:2006}. Recent studies searching for neutral gas outflows at $z\sim1$ have also shown evidence for fast outflows, although the sample size is still relatively small \citep{Aditya:2018a, Aditya:2019}. While FLASH will not be sensitive to the low optical depths typically observed in fast outflows for the majority of sources, it will provide a sample of powerful radio galaxies with large gas reservoirs which can be targeted with deeper observations using, for example, MeerKAT.  

The large sample of associated absorbers detected by FLASH will also allow us to explore the neutral gas properties of source populations with different AGN fuelling mechanisms, in particular by exploring the difference in gas kinematics and detection rates in High-Excitation and Low-Excitation Radio Galaxies (HERGS/LERGS; see e.g. \citealt{Chandola:2017,Chandola:2020}).

\subsubsection{The evolution of high-powered radio galaxies}

From the radio luminosity function for active galaxies at intermediate redshifts \citep{Sadler:2007}, we estimate that the FLASH survey volume contains at least 10,000 radio galaxies with redshifts in the range $0.4<z<1$ and continuum flux densities above 50\,mJy. These objects allow us to search for \mbox{H\,{\sc i}} in the host galaxies of radio AGN at lookback times of 4--8\,Gyr, when powerful radio galaxies were at least 3--5 times more abundant than they are today \citep{Heckman:2014, Pracy:2016}. These high-powered (predominately FR-II) radio galaxies are believed to be associated with gas-rich galaxy mergers, and are expected to be much richer in cold gas than FR-Is \citep{Heckman:1986, Carilli:1998}, but large systematic studies of the cold gas in these powerful radio galaxies have not been possible with existing radio telescopes. As such, the FLASH survey can provide important new constraints on the redshift evolution of the H{\sc i} content of radio galaxies.


\subsubsection{The multi-phase gas in AGN}

From associated absorbers we can determine properties of \mbox{H\,{\sc i}} near the centres of active galaxies, likely physically close to the host galaxy AGN. The presence of an active supermassive black hole is known to have a variety of impacts on the evolutionary path of the host galaxy \citep{Kauffmann:2003,DiMatteo:2005,Cattaneo:2009,Fabian:2012,Kormendy:2013}, although there are still many unanswered questions about the nature of the relationship between the two and the balance between fuelling, feedback and quenching in obscured AGN \citep{Morganti:2018, Hickox:2018, Odea:2020}. When we combine information on the neutral gas from FLASH with multi-wavelength information on the total hydrogen content from X-ray absorption, the presence of dust from infrared emission, and the distribution of ionised gas from optical spectroscopy, we gain a comprehensive physical insight into the properties of the multi-phase gas in AGN as a whole. As part of FLASH, we will use our unprecedented sample of \mbox{H\,{\sc i}} associated absorbers to bring together a new picture across the electromagnetic spectrum of this gas and how it relates back to the overall properties of each host galaxy and its evolution. 

Early FLASH data has already revealed evidence in support of a physical relationship between the gas detected by \mbox{H\,{\sc i}} absorption and soft X-ray absorption \citep{Moss:2017, Glowacki:2017}, and is supported by other studies \citep{Vink:2006, Ostorero:2010, Ostorero:2017, Ursini:2019}. Alongside the new window on the radio sky provided by ASKAP, the eROSITA telescope was launched in July 2019 and will transform our view of the X-ray sky \citep{Merloni:2012}. To facilitate further investigation into multi-wavelength absorption tracers using these next-generation instruments, we have formed a collaboration between the FLASH team and the eROSITA-DE team (SEAFOG: Studies with eROSITA and ASKAP-FLASH of Obscured Galaxies) dedicated to bringing together all-sky information on gas absorption in AGN. SEAFOG will give us the first large-scale homogeneous dataset of \mbox{H\,{\sc i}} absorption and X-ray absorption (also extremely well-matched in sensitivity and angular resolution), enabling us to perform an unbiased census of the population of \mbox{H\,{\sc i}} and X-ray absorbers in order to definitively uncover the relationship between them. 

FLASH also provides us with the possibility to study time variability in the multi-phase medium of active galaxies. Previous studies towards the nearby radio galaxy PKS\,1718-649 have found variability in both radio continuum and X-ray absorption on the timescales of months \citep[][Moss et al. in prep]{Tingay:2015,Beuchert:2018}. Modelling of the continuum variability indicates that the most likely mechanism here is free-free absorption \citep{Macquart:2016}, which suggests that we may be able to further trace this via variability in the \mbox{H\,{\sc i}} absorption profile. This particular galaxy has very weak \mbox{H\,{\sc i}} absorption (\citealt{Veron-Cetty:1995}), meaning that it is difficult to correlate any possible \mbox{H\,{\sc i}} absorption variability with the known radio continuum or X-ray variability. 

However, we expect with the much larger population of associated absorbers to be provided by FLASH, we will be able to find highly suitable targets to advance these studies. With eROSITA expected to revisit the sky eight times during its five year operational period and the excellent GHz continuum follow-up capabilities of the Australia Telescope Compact Array, we see a unique opportunity to conduct contemporaneous observational follow-up of early FLASH \mbox{H\,{\sc i}} absorption detections to search for correlated variability in radio continuum, hydrogen absorption and X-ray absorption simultaneously.


\subsection{Detecting molecular lines in the distant Universe} 

Molecular absorption lines provide an excellent probe of the physical and chemical conditions of the gas and its potential for star formation (e.g. \citealt{Israel:1988, Aalto:1995, Henkel:2005}). Furthermore, the relative line strengths can offer extremely accurate measures of the cosmic evolution of the microwave background temperature (\citealt{Wiklind:1997, Muller:2013}), time delays between sight lines can be used to measure of cosmological parameters (\citealt{Wiklind:2001}) and shifts in the line frequencies can measure the values of the fundamental constants at large look-back times (\citealt{Drinkwater:1998, Darling:2003, Curran:2004a}).

However, redshifted ($z>0.05$) millimetre (e.g. CO, HCN, HCO$^+$) and decimetre (OH) band absorption is currently very rare, detected in only six systems -- three intervening and three associated.\footnote{\citet{Wiklind:1995, Wiklind:1996a, Wiklind:1996b, Wiklind:1997} and \citet{Allison:2019} for the mm-band and \citet{Chengalur:1999, Kanekar:2002, Kanekar:2003, Kanekar:2005} and \citet{Gupta:2018} for the OH absorbers, four which are in common with the mm-band systems.}  Since DLAs exhibit a high column density of neutral gas at a known redshift, these have been the natural target for molecular line searches and, while CO has been detected (e.g. \citealt{Srianand:2008}) in the Lyman and Werner ultra-violet bands of the $\approx 60$ DLAs detected in \hbox{${\rm H}_2$} absorption (e.g. \citealt{Levshakov:1985, Balashev:2014, Noterdaeme:2015, Balashev:2019, Ranjan:2020}), millimetre and decimetre band searches have proven fruitless (\citealt{Curran:2004b, Curran:2006, Curran:2008, Curran:2011b, Kanekar:2014b, Gupta:2021}).\footnote{While there are over 150 detections of millimetre-wave band molecular emission at $z \gtrsim 1$ (e.g. \citealt{Omont:1996, Genzel:2015}), these are not coincident with 21-cm absorption (\citealt{Curran:2016c}). This could be due to more effective self shielding of the molecular gas (e.g. \citealt{Krumholz:2008, Krumholz:2009b, McKee:2010}) or the emission being from gas remote from the host galaxy, where the  \mbox{H\,{\sc i}} is expected to reside (\citealt{Papadopoulos:2000, Klamer:2004, Ivison:2012, Nesvadba:2009, Curran:2010}).}


The paucity of detections is most likely due to the low molecular column-density-fractions in DLAs, which, where detected, are $\sim10^{-7} - 0.3$ (\citealt{Molaro:2000, Noterdaeme:2017}, respectively), the majority of which are well below the limitations of pre-ALMA instruments (\citealt{Curran:2004b}).\footnote{The mean molecular fraction is $7.3\times10^{-4}$ and the median 0.0015.}
For the mm-band and OH absorbers the molecular fractions are $\approx0.6 - 1$ and the optical--near-infrared colours $V-K \gtrsim 5$ (\citealt{Curran:2006}), compared to $V-K \lesssim 4$ for the \hbox{${\rm H}_2$} bearing DLAs (\citealt{Curran:2011b}). This is a strong indication that the reddening is due to the presence of dust, which protects the molecular gas against the ambient UV field, and that future molecular absorption searches should therefore target the faintest optical sources. Another possible contributor in the low detection rate of mm-band absorption is that the cross-section of the molecular gas is very much smaller than that of the \mbox{H\,{\sc i}} (\citealt{Zwaan:2006}, but see also \citealt{Balashev:2018} for low-$N_{\rm{H_{2}}}$ gas).

It has been noted that the \mbox{H\,{\sc i}} 21-cm absorption strength also exhibits a correlation with both $V-K$ (\citealt{Curran:2019b}) and $B-K$ (\citealt{Curran:2017}) colours\footnote{Thereby, consistent with the higher detection rates along reddened sight lines (\citealt{Carilli:1998}).} and so the presence of 21-cm absorption can provide a valuable signpost to molecular absorption without the optical/UV bias towards less obscured sight lines. In fact, sources sufficiently obscured may consist predominately of molecular gas, where the atomic gas column density exceeds $N_{\rm HI} \sim 10^{22} \mathrm{cm}^{-2}$ (\citealt{Schaye:2001}).


While the mm-band lines provide very useful ``anchor'' lines with which to compare the shift in the observed 21-cm frequency, which is particularly sensitive to the fine structure constant (e.g. \citealt{Tzanavaris:2005}), the OH radical is particularly valuable as the four hyperfine lines at 18-cm have different dependences on various combinations the constants (fine structure, electron--proton mass ratio, proton g-factor), while being known to arise along the same sight-line (\citealt{Darling:2003, Chengalur:2003}). Furthermore, although the numbers are small, the known OH 18-cm absorbers have a similar velocity width to the 21-cm absorption profiles, indicating that this also arises along the same sight-line (\citealt{Curran:2007}). This suggests that the OH may not be subject to the same confinement affecting the mm-band observations (\citealt{Zwaan:2006}) and so the targeting of sufficiently obscured sources could significantly increase the number of OH detections.\footnote{The OH/\mbox{H\,{\sc i}} column density ratio ranges from $4\times10^{-7}.(T_{\text{spin}}/T_{\text{ex}})$ for $V-K =  4.8$ to $4\times10^{-5}.(T_{\text{spin}}/T_{\text{ex}})$ for $V-K =  8.9$ \citep{Curran:2011a}.}


\subsection{Tracing galaxy mergers using OH megamasers} 
OH megamasers (OHMs) are luminous 18\,cm masers that arise in extreme starbursts triggered by major galaxy mergers (e.g. \citealt{Darling:2007}). They are detectable at cosmological distances and can be used to measure magnetic fields via Zeeman splitting (\citealt{Robishaw:2008, McBride:2013}). The OH main lines are typically detected with the 1667~MHz line dominant and a non-thermal 1667:1665 MHz line ratio. Roughly 110 OHMs are known, up to a redshift of $z=0.265$ (\citealt{Roberts:2021, Baan:1992}). We expect FLASH to make direct detections of the redshifted 1667\,MHz line from megamaser galaxies at $0.67<z<1.34$. It may also be possible to detect OH megamasers by stacking the spectra of FIR luminous galaxies with available redshifts.

Since FLASH can detect OHMs at $0.67 < z < 1.34$ but is a shallow survey, it can only detect the extremal high-luminosity end of the OHM luminosity function.However, any OHM detected by FLASH will be the most distant (and likely the most luminous) known to date. Using the \citet{Roberts:2021} luminosity function and the areal and redshift coverage of FLASH, one can predict the number of OHMs detected by the survey, but the forecast is extremely sensitive to assumptions about the maximum possible OHM luminosity and the redshift evolution of major gas-rich galaxy mergers.

We assume an OH line width of 150 km~s$^{-1}$ and a line search of FLASH cubes that have been smoothed to match the associated lowered per-channel noise.  Assuming that the maximum isotropic OH luminosity detected will be twice the highest known luminosity ($L_{\rm max} = 2 \times 10^4$~$L_\odot$;  \citealt{Darling:2002}) and that the galaxy merger rate evolves as $(1+z)^\gamma$ where $\gamma = 2.2$ \citep{Roberts:2021}, we predict that $\sim$400 OHMs will be detected in FLASH in $0.67 < z < 1.09$.  This would grow the known sample by roughly a factor of four.

The primary uncertainty in this prediction lies in the unknown upper limit to the OHM luminosity.  The range $L_{\rm max} = 2 - 10 \times 10^4$~$L_\odot$ produces an order of magnitude change in the expected number of detections, which highlights the sensitivity of FLASH to only the most luminous OHMs. Given the large uncertainties in expected OHM detections, the outcome of a FLASH OHM survey will provide new insight into the OH luminosity function (and the associated maser physics, such as whether the higher gas fraction in galaxies at earlier epochs influences OHM luminosity and production rate) and the galaxy merger evolution.  OHMs likely arise during a specific late stage in the merger sequence and are therefore a probe of merging evolution that is less definition-, wavelength-, and resolution-dependent than other methods of assessing the merging history of galaxies. 
 

\subsection{Probing ionized gas and magnetic fields in 21-cm absorption galaxies} 

\mbox{H\,{\sc i}} absorption indicates neutral gas in intervening or associated absorbers, and provides the redshift and other properties of this gas. At the same time, if the illuminating continuum source is linearly polarized, this radiation will experience Faraday rotation, thereby tracing ionized gas and line-of-sight magnetic field components in the same absorbing systems.

Specifically, Faraday rotation measures the amount of rotation that linearly polarized background light experiences when passing through a magneto-ionic medium. This effect can be observed in linearly polarized Stokes parameters over large observing bandwidths. Quantifying polarized behaviour vs. $\lambda^2$ (the observing wavelength squared) can be used to infer the magnetic fields' strength, coherence, and turbulence. The Rotation Measure (RM) is commonly used to quantify Faraday rotation. Although the RM does not capture the entire broadband magnetic picture, it does provide a direct method to relate Faraday rotation to a line-of-sight magnetic field in the following manner:

\begin{equation}\label{equation:RM_def}
RM = 0.81 \int_{0}^{z_{\rm s}} n_{\rm e}(z) B_{\parallel}(z) (1+z)^{-2} \mathrm{d}l (z) \; \mathrm{rad \; m^{-2}},
\end{equation}

where $z_{\rm s}$ is the redshift of the emitting linearly polarized source, $n_{\rm e}$ is the column density of free electrons (in cm$^{-3}$) and $B_{\parallel}$  the line-of-sight magnetic field strength (in G), both measured along the line-of-sight ($\mathrm{d}l(z)$, in pc) to $z_{\rm s}$ (\citealt{Ferriere:2021}). \autoref{equation:RM_def} shows that Faraday rotation is measured as an integrated quantity along the line-of-sight; therefore, any contributions from foreground contaminants (such as the Milky Way or other intervening absorbers along the line-of-sight) must be subtracted off. Additionally,  because Faraday rotation traces both ionized gas ($n_{\rm e}$) and magnetic fields along the same sight-line,  it can allow us to estimate the gas phases in the absorbers.

In earlier magnetism studies, large bandwidth observations were not possible, and a linear fit between polarized angle vs. $\lambda^2$ for at least two observing bands were calculated, resulting in a single value RM (such techniques were used in \citealt{Taylor:2009}, currently still the largest RM catalogue to date). However, with larger bandwidth polarized observations, more robust and sophisticated methods (such as RM-synthesis and QU-fitting, \citealt{Brentjens:2005, OSullivan:2012}) have been developed to quantify Faraday rotation behaviour more accurately.

The FLASH Pilot survey will have substantial sky overlap with ASKAP's Polarization Sky Survey of the Universe's Magnetism (POSSUM) (\citealt{Gaensler:2010}) - providing broadband Stokes I, Q, U, and V information. In addition, POSSUM will have information quantifying polarization features using these more sophisticated methods, and hence properties of magnetic fields and ionized gas, of all polarized FLASH targets.  Specifically, POSSUM will produce a catalogue containing properties such as: the RM, width of the RM in Faraday space ($\sigma_{RM}$, and can characterize depolarization, \citealt{Sokoloff:1998}), the number of Faraday components (an output of RM-synthesis), 3D Faraday cubes to perform Faraday tomography (\citealt{Ideguchi:2018}), Faraday complexity information (e.g. \citealt{Anderson:2015}), and fractional linear polarization for all polarized sources within the survey field.

These polarized data will the study of the magnetic properties for both associated and intervening 21-cm absorbers while also providing information about the systems' ionized and neutral gas fractions. We will discuss two different statistical experiments of combining polarized observations and FLASH with both intervening and associated 21-cm absorption sources below. For these applications, AGN sight lines in which 21-cm absorption is not detected can be used as a statistical control sample to compare with associated and intervening 21-cm absorbers' magnetic properties.

\subsubsection{Intervening 21-cm absorbers}

Coherent $\mu$G magnetic fields in present day galaxies are thought to be the result of a large-scale dynamo, which orders and sustains fields via turbulence driven by supernova explosions or cosmic ray pressure and galactic differential rotation (\citealt{Ferriere:2000, Hanasz:2009}). The seed field for the large-scale dynamo could be a weak pre-galactic field, or an already-amplified $\mu$G field from a small-scale dynamo during the early phases of galaxy formation (e.g. \citealt{Kronberg:1999, Furlanetto:2001, Arshakian:2009, Rieder:2016}). Testing these different theoretical models remains challenging due to the few galactic magnetic field measurements we have beyond the local Universe (e.g. \citealt{Oren:1995, Bernet:2008, Farnes:2014, Kim:2016, Mao:2017}).

To overcome the intrinsic faintness of the polarized synchrotron emission from distant galaxies, we need to use absorption-type experiments - distant galaxies seen in projection against polarized background radio sources - to characterize their magnetic fields. Coherent magnetic fields in the intervening galaxy could produce a net Faraday rotation signal while turbulence in the intervening galaxy could depolarize the background radiation and induce Faraday complexities. Synergies between the FLASH and POSSUM surveys will offer an unique opportunity to study magnetic fields in intervening \mbox{H\,{\sc i}} absorption galaxies in the redshift range $0.4 < z < 1.0$. For FLASH detected 21-cm intervening absorbers with a polarized background continuum source, POSSUM will provide information on the Faraday rotation and depolarization properties for the sight line. This will yield, for the first time, a statistically significant sample of few 10s to a few hundred 21-cm intervening absorbers with polarization information. With a carefully selected sample of control sight lines towards background polarized sources without any intervening 21-cm absorption, we will statistically infer the typical Faraday rotation and depolarization produced by the intervening galaxy population (\citealt{Basu:2018}). When possible, we will further convert these observables into magnetic field strength estimates using the FLASH \mbox{H\,{\sc i}} column density constraints and an assumed ionization fraction. As stated in \autoref{section:nature_absorption-selected_galaxies}, many host galaxies of FLASH 21-cm absorbers are expected to be identified/characterized, therefore, enabling one to directly link the derived magnetic field properties to physical properties of the absorber galaxies population. This has recently been demonstrated using FLASH commissioning data where \mbox{H\,{\sc i}} absorption was detected towards the radio lobe of the powerful radio galaxy PKS\,0409$-$75, allowing the magnetic field of the absorbing galaxy at $z=0.67$ to be estimated (\citealt{Mahony:2021}). 

In addition, FLASH itself, with its full-Stokes data cube, could detect Zeeman splitting of 21-cm absorption line and provide in-situ galactic magnetic field measurements for 21cm intervening absorbers at intermediate redshifts $0.4 < z < 1.0$. An upper limit of line-of-sight B field of 17 $\mu$G was placed towards an 21-cm absorber at $z = 0.692$ (\citealt{Wolfe:2011}). Sufficiently long integration towards DLA systems could yield an interesting sample of in-situ magnetic field estimates out to $z \sim$ 1 galaxies and further constrain the time-scale of galactic magnetic field generation. 

\subsubsection{Associated 21-cm absorbers}

The magnetic properties of \mbox{H\,{\sc i}} 21-cm associated absorbers have never been previously studied. For associated absorbers, the Faraday rotation measurements come from within the AGN itself. With combined information from FLASH and polarized observations (like POSSUM), we can address the interplay between cold neutral and ionized gas in AGN and their magnetic fields in the immediate environment of the AGN. However, the magnetic fields within AGN are still not well understood \citep{Agudo:2015, Beck:2009}. Moreover, we do not understand the relationship between how Faraday rotation behaviour, and thus magnetic fields, play into AGN morphology \citep{Anderson:2015, OSullivan:2017}.

Few polarization studies have been done on 21-cm absorption sources. CSS sources are better studied, and many are detected with \mbox{H\,{\sc i}} absorption (\autoref{section:observations_associated_absorbers}). \cite{Saikia:2003} conducted a radio polarization study using young resolved CSS sources from literature (without confirmed 21-cm absorption) and found asymmetric polarization properties between the lobes of these sources. \cite{Teng:2013} analyzed the polarized spectra of 27 local ($z < 0.12$) AGN with \mbox{H\,{\sc i}} absorption. They found that \mbox{H\,{\sc i}} absorption features were polarization-dependent.  Both studies suggest the polarization behaviour of these sources could be explained by interactions between the radio jets and lobes and in-falling or out-flowing material.

In \autoref{section:coevolution_agn} we discuss how associated 21-cm absorbers can be used to study the interaction between neutral gas and the AGN. A large amount of cool gas content could cause increasing accretion rates, which could affect the magnetic fields. Higher accretion rates can cause stronger collimated jets \citep{Allen:2006} containing ordered or helical magnetic fields \citep{Gabuzda:2018}. In addition to gas triggering accretion, previous studies have shown that associated 21-cm absorbers have been associated with high-velocity outflows, possibly through large radio jets \citep{Morganti:2015, Gupta:2016}. The mixing of large amounts of gas and magnetic fields can cause turbulent magnetic fields and depolarization. \cite{OSullivan:2015} showed that radiative-mode AGN also contained smaller fractional polarization, indicating depolarization attributed to the AGN's gaseous environment. 

There are 82 known associated 21-cm associated absorbers at $z > 0.1$ \citep{Curran:2021}, most of which have been found through optical spectra (as discussed in \autoref{section:observations_associated_absorbers}) with no radio polarization follow-up information to perform magnetism studies. However, with a sizeable polarized sample of associated \mbox{H\,{\sc i}} absorbers from the combination of FLASH and POSSUM, statistical analysis can be done to determine these sources' general magnetic properties and provide insight into the role of neutral gas and magnetic fields in the immediate AGN environment.

Specifically, FLASH and POSSUM can tell us whether associated 21cm absorption sightlines contain enhanced Faraday rotation, increased depolarization, and increased Faraday complexity compared to a control sample.  Such properties can indicate stronger magnetic fields and mixing between magnetic fields and the surrounding gas, relating to outflows and environmental factors within associated absorbers. To confirm the role of such outflows in the magnetic fields of associated absorbers, the neutral gas velocity of the associated absorbers can be calculated from the FLASH catalogue and compared with the magnetic properties mentioned above. To complement this study, pre-existing multi-wavelength data can provide information about the associated absorbers host galaxy's mass, luminosity, and metallicity (inferring star formation history). Meanwhile, Faraday rotation growth with FLASH-derived redshifts would indicate the evolution of AGN gas and magnetic fields through cosmic time. 
 
In addition to learning about magnetic fields and neutral gas in the immediate AGN environment, we can also learn about their ionization states. As mentioned in \autoref{section:hi-absorption_in_galaxies}, most associated absorbers have been detected for $z < 1$, with a critical ionizing luminosity as the proposed reason for this effect. \autoref{equation:RM_def} shows that the RM is directly proportional to the number of free electrons $n_{\rm e}$ along the line-of-sight. We can test whether Faraday rotation can also act as a proxy for the ionization state of systems, allowing allow us to explore the idea of a critical ionizing luminosity being responsible for the lack of 21-cm absorbers in the high redshift universe.


\subsection{The diffuse ionized gas in galaxies using RRL emission}

Diffuse ionized gas is another integral part of the ISM of galaxies and an area in which novel observations of radio recombination lines (RRLs) in FLASH could have a profound impact. By looking across populations of sources, FLASH will provide some of the first insights into the detectability of stimulated RRLs arising from diffuse ionized gas ($n_{\rm e} \sim 1 - 1000$~\cmc) out to $z \sim 2$. RRL detections in FLASH will (a) investigate the role of the ISM in impeding AGN jet activity in bright, peaked-spectrum sources, (b) investigate the heating of the ISM by photoionization in the star-forming galaxies of intervening systems, and (c) lead to an unprecedented view in the Milky Way Galaxy of diffuse ionized gas in star-forming regions, at the crucial frequencies where RRL emission from this gas is brightest. 

In gas with low density, RRL transitions may be stimulated by radio continuum emission \citep{Shaver:1975a}. In the low-optical-depth regime, the stimulated RRL intensity is proportional to the continuum intensity, $S_{\mathrm{ RRL}} \approx \tau S_{\rm c}$, enabling RRLs to be observed in and against bright continuum sources out to high redshift. When the intensity of stimulated RRLs is measured at different frequencies (or equivalently, principal quantum numbers of the RRL transitions), the electron temperature, electron density, and pathlength of emitting regions can be precisely determined \citep[within 15\%;][]{Shaver:1975a, Oonk:2017}. 

The time is now ripe for FLASH to make the first statistical census of RRLs from diffuse ionized gas at cosmological distances. The very first extragalactic detections of RRLs, in M82 and NGC253, found contributions from stimulation \citep[e.g.,][]{Shaver:1978b}. Quickly after, \citet{Churchwell:1979} used the Arecibo 300~m telescope to search 21 galaxies and AGN with the 1.4 GHz receiver and 3 AGN with the 430 MHz receiver, but did not detect emission of $\Delta{S}/S_{\rm c} \gtrsim 10^{-3}$. To similar sensitivities, \citet{Bell:1984} used the Effelsberg 100~m dish at 4.8~GHz to search 10 galaxies without clear success. Yet these observations were carried out using narrow bandwidths that also depended on redshifts to sources which have since become more accurate. On the other hand, \cite{Bell:1980} detected the H83$\alpha$ and H99$\alpha$ lines at 10.5 GHz and 6.2 GHz, respectively, associated with the peaked-spectrum source OQ~208 at $z = 0.0763$, demonstrating a clear capability outside of the local universe. To date, 7 of the 21 galaxies with detected RRL emission show evidence for stimulated emission by non-thermal emission \citep[for an overview, see][]{Emig:2021}. Recently, \citet{Emig:2019} detected stimulated RRLs, with a rest-frame frequency of 284~MHz, in an intervening galaxy at $z = 1.124$ along the line of sight to 3C~190. The improved sensitivity of high-resolution interferometers and the large fractional bandwidths that enable deeper searches through line stacking are making RRL observations now feasible. The development of new cross-correlation techniques enables blind searches of RRLs across redshift space \citep{Emig:2020a}. 

\begin{figure}
    \centering
    \includegraphics[width=\columnwidth]{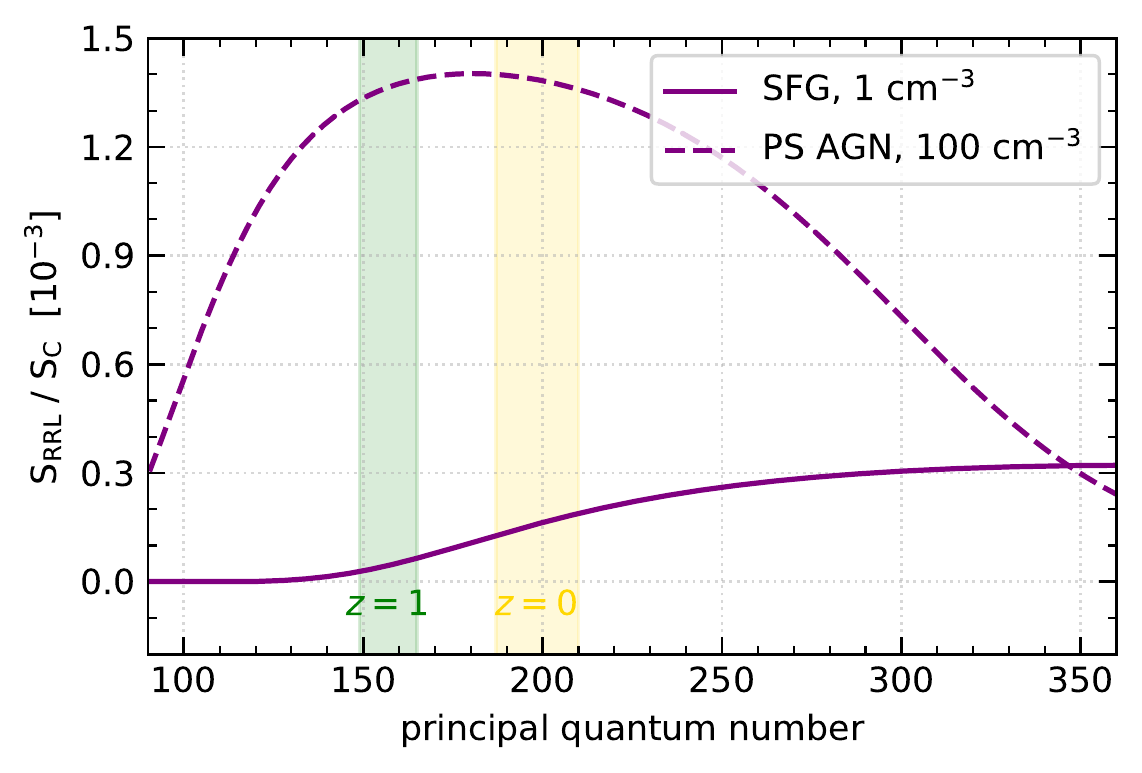}
    \caption{Line to continuum ratio for stimulated hydrogen RRL emission for two representative sources that FLASH will investigate. We plot the expected fractional emission, with a solid line, for an intervening star-forming galaxy (SFG) represented with $n_{\rm e} =1$~\cmc\ and $EM = 10^3$~\emuni, and with a dashed line, a peaked-spectrum AGN (PS AGN) with $n_{\rm e} =100$~\cmc\ and $EM = 10^5$~\emuni. The shaded green (yellow) region indicates the principal quantum numbers covered by FLASH for sources at $z = 1$ ($z = 0$). }
    \label{fig:rrl}
\end{figure}

In diffuse ionized gas, hydrogen emits the brightest in RRL emission --- by a factor of 5 or more than helium, and 10\,000 more than carbon. In Figure~\ref{fig:rrl}, we illustrate expected hydrogen RRL emission in terms of a fraction of the continuum flux density. One case shows the expectation for a star-forming galaxy, that may be intervening along the line-of-sight to a radio-bright AGN or a nearby bright galaxy. We consider an effective emission measure of $EM = 1000$~\emuni\ and density $n_{\rm e} = 1$~\cmc, representing a pervasive ionized component with somewhat elevated thermal pressures that reflect  conditions in starburst galaxies. The second case we show is for diffuse gas that may be present intrinsic to the ISM of a peaked-spectrum source, with $n_{\rm e} = 100$~\cmc\ and $EM = 10^5$~\emuni, an emission measure which corresponds to optically thick free-free absorption ($\tau_{\mathrm{ff}}= 1$) at 202~MHz \citep{Condon:1992}. The peak line to continuum ratio for RRLs is expected to be $\sim 10^{-4} - 10^{-3}$ or less. In the computations of Figure~\ref{fig:rrl}, we follow the procedure of \citet[][see their Equation 2]{Emig:2019} for a single slab of gas and assume stimulation by non-thermal emission only, an electron temperature of $10^4$~K, a line width of 150~\kms\ and incorporate line broadening terms as outlined by \citet{Salas:2017}. We use the models of \citet{Salgado:2017a} for non-LTE coefficients.

\begin{table}
    \centering
    \caption{Optical depth sensitivity of radio recombination line emission for a stacked population of intervening absorbers, $\left< \tau_{\rm{RRL,\,total}}^{\rm int} \right>$, and associated absorbers, $\left< \tau_{\rm{RRL,\,total}}^{\rm asc} \right>$ (see \autoref{table:nabs}).}
    \begin{tabular}{lccc}
    \hline \hline
    $T_{\rm s}$ & 100\,[K] & 300\,[K] & 1000\,[K] \\
    \hline
    \multicolumn{4}{c}{\bf Intervening absorbers} \\
    $\mathcal{N}_{\rm abs}^{\rm int}$ & 2800 & 850 & 180 \\
    $\left< S_{\rm c} \right>$ & 600\, [mJy] & 1381\, [mJy] & 4338\, [mJy] \\
    $\left< \tau_{\rm{RRL,\,total}}^{\rm int} \right>$ & 4.4 $\times 10^{-5}$ & $3.5 \times 10^{-5}$ & 2.4 $\times 10^{-5}$ \\
    \hline
    \multicolumn{4}{c}{\bf Associated absorbers} \\
    $\mathcal{N}_{\rm abs}^{\rm asc}$ & & 2000 & \\
    $\left< S_{\rm c} \right>$ & & 530\,[mJy] & \\
    $\left< \tau_{\rm{RRL,\,total}}^{\rm asc} \right>$ & & $5.9 \times 10^{-5}$ & \\
    \hline \hline 
    \end{tabular}
    \label{tab:rrl_est}
\end{table}

The frequency coverage of FLASH between 711.5 MHz and 999.5 MHz includes 23 (18) transitions from principal quantum numbers $\mathsf{n} = 187 - 209$~$(149 - 166)$ at a redshift of $z = 0$ ($z = 1$). FLASH can detect a continuous range of RRL redshifts, including redshifts beyond the \mbox{H\,{\sc i}} 21 cm redshift limit, as long as a radio source is background to the gas. The brightest AGN with known redshifts will enable the deepest searches intrinsic to the systems. We will also search for RRLs in diffuse gas at the redshifts of known absorbers: 1) known and discovered through \mbox{H\,{\sc i}} absorption, and 2) absorption line systems known through optical or other means, which extend to any redshift that is less than the target AGN. The FLASH survey will lay the ground-work for follow-up multifrequency observations which could then be used to determine gas physical properties and investigate the detailed physics needed to understand (a) the evolution and duty cycle of AGN and (b) the heating and structural conditions of the ISM.

We provide estimates on the sensitivity of FLASH to RRL optical depth. In our calculations, we conservatively assume 13 lines will be available to stack in each source. Following the predictions described in Section 3.3, we compute the average continuum flux density, $\left<S_{\rm c}\right>$, at 844 MHz as a weighted average over the expected number of intervening or associated H~I absorbers, see Table~\ref{tab:rrl_est}. Stacking RRLs at the redshift of all intervening sources, FLASH will probe deep RRL optical depths of $\tau_{\rm RRL} \approx (2-4) \times 10^{-5}$, to reach a peak signal-to-noise ratio of 3 for 36~\kms\ channel widths, see Table~\ref{tab:rrl_est}. For the star-forming galaxy (SFG) example with $n_{\rm e} = 1$~\cmc\ presented in Figure~\ref{fig:rrl}, an RRL optical depth of $\tau_{\rm RRL} = 3.5 \times 10^{-5}$ reaches emissions measures of $EM = 350$~\emuni. Similarly, for associated gas, RRL optical depths of $\tau_{\rm RRL} \approx 5.9 \times 10^{-5}$, at a peak signal-to-noise ratio of 3 for 36~\kms\ channel widths, will be reached. For the peaked spectrum AGN (PS AGN) diffuse gas model with $n_{\rm e} = 100$~\cmc, an RRL optical depth of $\tau_{\rm RRL} = 5.9 \times 10^{-5}$ reaches emissions measures of $EM = 4200$~\emuni.

Within the Galaxy, the large bandwidth and unprecedented frequency coverage of FLASH will probe a key but previously missing frequency range in studying diffuse ionized gas within the Galactic plane. With sensitivity to largest angular scales of 21\amin, hydrogen RRLs observations will probe the low density ionized gas towards the Inner Galaxy and in leaky \Hii\ regions, photoevaporating surfaces also referred to as ionized boundary layers, and the diffuse ionized component often surrounding embedded \Hii\ regions \citep[e.g.,][]{Anantharamaiah:1986, Goldsmith:2015}. FLASH observations used in conjunction with existing RRL observations at higher and lower frequencies and similar spatial resolution \citep[e.g.,][]{Alves:2015, Anderson:2021} will be used to measure how the RRL intensity changes as a function of frequency (as shown in Figure~\ref{fig:rrl}) and determine the gas density and temperature (Salas et al. in prep.). Helium and carbon RRLs will also be covered by the observing set-up. Where present, helium RRLs will probe the metallicity of diffuse ionized gas. Carbon RRLs may also be detected at FLASH frequencies but would arise from colder and denser conditions of the ISM (e.g., Roshi et al., submitted).


\section{Modelling and simulations}\label{section:simulations}

Realistic survey simulations are an important tool for interpreting data from the FLASH survey, as they allow us to link the observations to the underlying physical processes governing the distribution and evolution of H{\sc i} gas. 

The simulation requirements of an all-sky H{\sc i} absorption survey like FLASH are unique. Interpreting the survey results requires knowledge of the multi-phase ISM and in particular the separation between the cold/warm neutral medium and the molecular component, in addition to large-number statistics of galaxy properties. The latter is perhaps the easiest to reconcile; current cosmological-scale galaxy formation simulations are able to reproduce a large range of global observational relations, including galaxy luminosity, mass and clustering functions. Moreover, using post-processing methods, these simulations are largely able to reproduce global H{\sc i} relations, such as the H{\sc i} -- M$_{\star}$ relation \citep[e.g.][]{Crain:2017, Diemer:2019} and the H{\sc i} mass -- size relation \citep{Bahe:2016}. For a recent review on the cold and molecular gas properties of galaxies in recent cosmological hydrodynamic simulations, see \citet{Dave:2020}.

FLASH will use the results from two types of cosmological-scale simulations; `semi-analytic' models (SAMs) and hydrodynamic simulations. The former use dark matter halo merger histories derived from the outputs of cosmological N-body simulations, and follow the formation and evolution of galaxies using a set of physically-motivated prescriptions. These account for galaxy evolutionary processes such as: gas accretion, AGN feedback, stellar feedback, cooling, disk formation and star formation, along with their inter-dependencies. For a review on the physical processes that shape galaxy properties, see \citet{Somerville:2015}. SAMs are a computationally cheap, powerful tool, allowing us to model the galaxy distributions in the large cosmic volumes corresponding to ASKAP survey sizes.

On the other hand, hydrodynamic simulations use the equations of fluid dynamics to follow gas on top of the N-body component (representing dark matter/stars). The most popular numerical methods for solving the fluid dynamic equations discretise gas using point masses (as in Smoothed Particle Hydrodynamics, or SPH; \citealt{Gingold:1977}, \citealt{Lucy:1977}), a static, adaptive grid/mesh (as in Adaptive Mesh Refinement, or AMR; \citealt{Berger:1984}, \citealt{Berger:1989}), or using a moving mesh (e.g. \citealt{Springel:2010}). These simulations have the advantage that they can self-consistently follow processes such as the mixing of gas phases, feedback-generated shocks, and the growth of fluid instabilities. However, they are more computationally intensive than SAMs, and therefore must focus on smaller cosmological volumes. Moreover, the physical processes that can be investigated are highly dependent on the mass/spatial resolution of the simulation. 

FLASH requires a holistic approach to simulations; SAMs/large-scale hydrodynamic simulations will be used to investigate the distribution of background radio galaxies along with their properties, while smaller-scale, very high resolution hydrodynamic simulations are required to understand the detailed physical state of the absorbers within intervening galaxies. These can be in the form of idealised, non-cosmological, simulations of galaxies, or cosmological zoom-in simulations, which simulate a region of a larger cosmological-scale simulation at a significantly higher resolution, in an effort to consider both the detailed multi-phase ISM of a galaxy, alongside environmental processes such as its merger/assembly history. 

In their recent paper, \citet{Garratt-Smithson:2021} used the EAGLE suite of cosmological simulations \citep{Schaye:2015} in order to explore the nature of high column density \mbox{H\,{\sc i}} absorbers in the redshift range of interest to FLASH. The authors noted a redshift evolution in the DLA covering fraction -- $M_{\star}$ relation, along with a dependence of DLA properties on galaxy evolutionary processes such as AGN feedback. Moreover, the authors found a significant mass of \mbox{H\,{\sc i}} in the outskirts of galaxies and the circumgalactic medium (CGM). This fits with the recent early science results from FLASH \citep{Allison:2020}, where the authors detected absorption in the outskirts (impact parameter $= 17\,$\,kpc) of an intervening early-type galaxy.  This result also links to recent theoretical work by \citet{Nelson:2020}, who predict tens of thousands of discrete absorption systems, with sizes of order a kpc or smaller, in the CGM of elliptical galaxies at intermediate redshift ($\sim 0.5$).


\section{Data products}\label{section:data_products}

ASKAP data products will be generated through a dedicated pipeline using the \textsc{ASKAPsoft} package (e.g. \citealt{Cornwell:2011, Guzman:2019, Wieringa:2020}). Quality assessment will be carried out by the FLASH survey science team for each pointing, after which data products will be publicly released via the CSIRO ASKAP Data Archive (CASDA)\footnote{\url{https://research.csiro.au/casda/}}. These will include fully-mosaicked, primary-beam-corrected continuum and spectral-line images, spanning the full bandwidth and field of view. 

The FLASH continuum images will complement the shallower Rapid ASKAP Continuum Survey \citep[RACS;][]{McConnell:2020} and deeper Evolutionary Map of the Universe (EMU; \citealt{Norris:2011, Norris:2021}), by sitting between them in terms of sensitivity and integration time. Based on the results of the recently completed RACS, we expect a median image noise in the FLASH continuum images of approximate 93\,$\mu$Jy\,beam$^{-1}$, which is about a factor of 3 more sensitive than RACS.

These continuum images will then be used to generate a catalogue of sources using the \textsc{Selavy} image analysis tool (\citealt{Whiting:2012}) that is built into the pipeline. The brightest source components ($S_{\rm 850\,MHz} \gtrsim 10$\,mJy) will then be used to generate one-dimensional spectra from the spectral-line cubes to search for absorption lines. These spectra will include estimates of the spectral-line flux density, rms noise per spectral channel and background continuum flux density.

In addition to the standard data products produced by the pipeline, the FLASH survey science team plans to release value-added catalogues that will include multi-wavelength information on both the radio sources and detected 21-cm absorbers. Following publication, these value-added catalogues will be publicly accessible via CASDA and/or a dedicated consolidated \mbox{H\,{\sc i}} absorption data base.


\section{Summary}\label{section:summary}

FLASH is the survey for extragalactic \mbox{H\,{\sc i}} 21-cm absorption using the ASKAP radio telescope. It will cover the entire sky south of $\delta \approx +40 \deg$ at frequencies covering $\nu = 711.5 - 999.5$\,MHz, thus including $\mbox{H\,{\sc i}}$ redshifts between $z = 0.4$ and $1.0$. The expected spectral rms noise is 3.2 -- 5.1\,mJy\,beam$^{-1}$ per 18.5\,kHz, and the continuum rms noise will be 93\,$\mu$Jy\,beam$^{-1}$. 

The total absorption path length sensitive to DLA-like column densities will be $\Delta{X} \sim 10^{4}$; two orders of magnitude greater than previous surveys. FLASH will yield more than a thousand intervening and intrinsic/associated 21-cm absorbers, enabling a homogeneous flux-limited survey of the cold \mbox{H\,{\sc i}} content of galaxies at intermediate cosmological distances. Likewise, the wide-field continuum images will provide the most sensitive census of radio AGN over the entire sky until the deeper EMU survey is available.

The key science goals are to determine how the cool \mbox{H\,{\sc i}} gas in galaxies has evolved since cosmic noon and examining the gas accretion mechanisms that drive the co-evolution of supermassive black holes and their host galaxies over cosmic history. In addition to these, we  expect to detect rare molecular absorption lines, OH emission associated with megamasers, and RRLs associated with diffuse ionized gas in the ISM. Finally, we expect that the spectra and continuum maps generated from FLASH will provide a valuable legacy resource for astronomers for the next decade.   


\begin{acknowledgements}
JRA acknowledges support from a Christ Church Career Development Fellowship. Parts of this research were conducted by the Australian Research Council Centre of Excellence for All-sky Astrophysics in 3D (ASTRO 3D) through project number CE170100013. TA acknowledges support from  National Key R\&D Programme of China (2018YFA0404603). ACE acknowledges support from STFC grant ST/P00541/1. MG was partially supported by the Australian Government through the Australian Research Council's Discovery Projects funding scheme (DP210102103). ZZ is supported by NSFC grant No. 11988101, U1931110 and 12041302. ZZ is also supported by CAS Interdisciplinary Innovation Team (JCTD-2019-05).

The Australian SKA Pathfinder is part of the Australia Telescope National Facility which is managed by CSIRO. Operation of ASKAP is funded by the Australian Government with support from the National Collaborative Research Infrastructure Strategy. ASKAP uses the resources of the Pawsey Supercomputing Centre. Establishment of ASKAP, the Murchison Radio-astronomy Observatory and the Pawsey Supercomputing Centre are initiatives of the Australian Government, with support from the Government of Western Australia and the Science and Industry Endowment Fund. We acknowledge the Wajarri Yamatji people as the traditional owners of the Observatory site.

We have made use of \textsc{Astropy}, a community-developed core PYTHON package for astronomy (\citealt{Astropy:2013, Astropy:2018}), NASA's Astrophysics Data System Bibliographic Services, and the VizieR catalogue access tool operated at CDS, Strasbourg, France.
\end{acknowledgements}


\begin{appendix}

\section{Estimating survey outcomes}\label{section:estimating_survey_outcomes}

In this section we describe in more detail the methodology used to determine the expected outcomes that guided the FLASH survey design.


\subsection{Input source catalogue}\label{section:input_source_catalogue}

To estimate what we expect to detect with FLASH, we simulate a statistically-realistic sample of radio sources using the entire source catalogue of the National Radio Astronomy Observatory Very Large Array Sky Survey (NVSS, $\nu = 1.4$\,GHz, $S_{\rm src} \gtrsim 2.5$\,mJy; \citealt{Condon:1998}) and scale to the survey area as required. To avoid sources that are likely to be in the immediate foreground, we apply a maximum angular size filter of 60\,arcsec ($d_{\rm src} \sim 30$\,Mpc at $z \sim 1$). The source flux densities are estimated at each frequency $\nu$ by extrapolating from the catalogue values at $1.4$\,GHz and assuming a canonical spectral index of $\alpha = -0.7$. Each of these sources then forms a simulated sight-line for which we can determine the expected number of absorbers that we will detect with FLASH. 


\subsection{Completeness}\label{section:completeness}

\begin{figure}
\begin{center}
\includegraphics[width=\columnwidth]{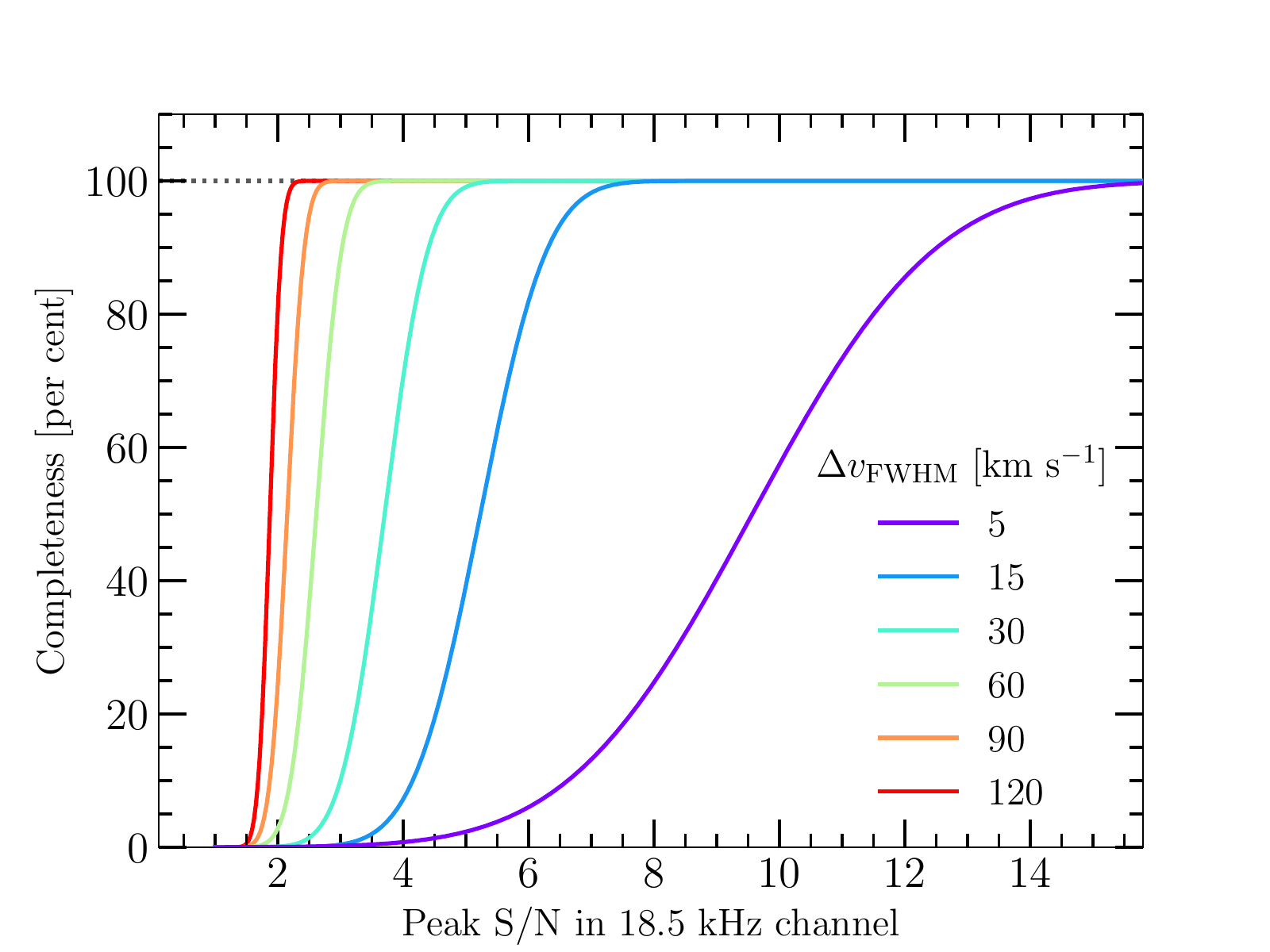}
\caption{The expected completeness of reliable detections in FLASH data as a function of the peak S/N in a single 18.5-kHz channel and FWHM. Based on completeness simulations by \citet{Allison:2020} for the FLASH Early Science survey of the GAMA\,23 field.}\label{figure:completeness}
\end{center}
\end{figure}

We define the completeness as the probability that an absorption line of given peak S/N and velocity width is recovered from the FLASH spectra using our detection method. To detect absorption lines we use a bespoke software tool called \textsc{FLASHfinder}\footnote{\url{https://github.com/drjamesallison/flash_finder}} (\citealt{Allison:2012b}), which uses the \textsc{PyMultiNest} (\citealt{Buchner:2014}) implementation of \textsc{MultiNest} (\citealt{Feroz:2008, Feroz:2009, Feroz:2013}) to calculate the Bayes factor $B$ (e.g. \citealt{Kass:1995}) for a simple Gaussian absorption line versus a null model. The assumption that a 21-cm absorption line approximates a Gaussian line profile (or superposition thereof) is due to the fact that broadening is dominated by Doppler shift, resulting from a sight line through a velocity field containing superposition of thermal, turbulent and rotational motion. 

To determine the expected reliability of detections, we use the early science ASKAP observations of the GAMA\,23 field by \cite{Allison:2020}, who found that a Bayes Factor $\ln{B} > 16$ was sufficient for a reliable detection of an absorption line. This was largely due to the effect of a non-Gaussian multiplicative component in the noise caused by a channelisation error in the correlator, resulting in a background level of false-positive detections. Although this error is now fixed, there is a strong likelihood that any blind survey for absorption will contain a background level of non-Gaussian and multiplicative noise (due to hardware and/or calibration and data processing errors) that will need to be accounted for. We therefore take a conservative approach and use the results of \cite{Allison:2020} as an indication of future results. The completeness was then determined by simulating 1000 absorption lines in bins of peak S/N and width and using the fraction that were recovered. The results are shown in \autoref{figure:completeness} and define the expected completeness as a bivariate function of peak S/N and FWHM. 


\subsection{Total absorption path length}\label{section:absorption_path_length}

The comoving absorption path length spanned by redshift interval $\mathrm{d}{z}$ is given by
\begin{equation}
 \mathrm{d}X = \mathrm{d}{z}\,(1+z)^{2}\,E(z)^{-1},
\end{equation}
where
\begin{equation}
\mathrm{d}z = (1+z)\frac{\mathrm{d}\nu}{\nu},
\end{equation}
\begin{equation}
z = \frac{\nu_{\rm HI}}{\nu} - 1,
\end{equation}
\begin{dmath}
 E(z) = {\sqrt{(1+z)^{3}\,\Omega_{\rm m} + (1+z)^{2}\,(1 - \Omega_{\rm m} - \Omega_{\Lambda}) + \Omega_{\Lambda}},}
\end{dmath} 
$\nu$ is the corrected observed frequency, and $\nu_{\rm HI}$ is the rest frequency of the \mbox{H\,{\sc i}} 21-cm line, equal to 1420.40575177\,MHz (\citealt{Hellwig:1970}).

The total comoving absorption path length for a column density sensitivity $N_{\rm HI}$ is then given by the following sum over sight lines towards $n$ sources
\begin{equation}\label{equation:total_absorption_path}
    \Delta{X}(N_{\rm HI}) = \sum_{i=1}^{n} \int_{z_{\mathrm {min}}}^{z_{\mathrm {max}}} {C_i(N_{\rm HI}, z)\,w_i(z + \Delta{z}_{\rm asc})\,\mathrm{d}{X(z)}},
\end{equation}
where $z_{\rm min}$ and $z_{\rm max}$ are the minimum and maximum redshift in the survey, $C_i( N_{\rm HI}, z)$ is the probability that an absorption line of column density $N_{\rm HI}$ is recovered at redshift $z$  (i.e. the completeness) and $w_i(z + \Delta{z}_{\rm asc})$ is the probability that the $i$'th source is located at a redshift greater than $z + \Delta{z}_{\rm asc}$. Absorption associated with \mbox{H\,{\sc i}} gas in the host galaxy of the radio source is excluded by using an offset in redshift, $\Delta{z}_{\rm asc}$, given by 
\begin{equation}
    \Delta{z}_{\rm asc} = (1 + z) \Delta{v}_{\rm asc}/c,
\end{equation}
where $\Delta{v}_{\rm asc} = 3000$\,km\,s$^{-1}$.

We determine $C_i(N_{\rm HI}, z)$ by applying the bivariate completeness function given in \autoref{section:completeness} to the expected peak S/N and FWHM. The expected peak S/N is given by (see \autoref{equation:optical_depth})
\begin{equation}
    \mathrm{S/N} = \left[{1 - \exp(-\tau)}\right] \frac{c_{\rm f}\,S_{\rm c}(z)}{\sigma_{\rm chan}(z)},
\end{equation}
where $S_{\rm c}$ is the source flux density, $c_{\rm f}$ is the source covering factor and $\sigma_{\rm chan}$ is the rms channel noise. For a Gaussian line profile with FWHM $\Delta{v}_{\rm FWHM}$, the peak optical depth is related to $N_{\rm HI}$ by (see \autoref{equation:column_density})
\begin{equation}
	\tau = 0.011 \left[\frac{N_{\rm HI}}{2 \times 10^{20}\,\mathrm{cm}^{-2}}\right] \left[\frac{T_{\rm s}}{300\,\mathrm{K}}\right]^{-1} \left[\frac{\Delta{v}_{\rm FWHM}}{30\,\mathrm{km}\,\mathrm{s}^{-1}}\right]^{-1},
\end{equation}
where $T_{\rm s}$ is the spin temperature. 

For $w_i(z)$\footnote{In the case of real survey data we will have multiwavelength spectroscopic and photometric information that will allow us to more precisely determine the redshift for some sources.} we use the redshift distribution determined by \cite{deZotti:2010} from a fit to bright ($S_{\rm 1.4} \geq 10$\,mJy) sources in the Combined EIS-NVSS Survey Of Radio Sources (CENSORS; \citealt{Brookes:2008}),
\begin{equation}\label{equation:redshift_weighting}
w_{i}(z) = \int_{z}^{\infty}{\mathcal{\hat{N}}_{\rm src}(z^{\prime})\,\mathrm{d}z^{\prime}},
\end{equation}
where the un-normalised redshift distribution is given by
\begin{equation}\label{unnormalised_source_redshifts}
\mathcal{N}_{\rm src}(z) = 1.29 + 32.37\,z - 32.89\,z^{2} \
 + 11.13\,z^{3} - 1.25\,z^{4},
\end{equation}
and 
\begin{equation}\label{equation:normalised_source_redshifts}
\mathcal{\hat{N}}_{\rm src}(z) = \frac{\mathcal{N}_{\rm src}(z)}{{\int_{0}^{\infty}{\mathcal{N}_{\rm src}(z^{\prime})\,\mathrm{d}z^{\prime}}}}.
\end{equation}
We assume that this redshift distribution applies to any sight line irrespective of the source flux density (e.g. \citealt{Condon:1984}), which is the result of a strongly evolving radio-loud AGN population.


\subsection{Number of intervening 21-cm absorbers}\label{section:no_intervening_absorbers}

We calculate the expected number of intervening 21-cm absorbers of a given column density $N_{\rm HI}$ or less by evaluating the following sum over $n$ sight lines
\begin{multline}\label{equation:intervening_absorbers}
	\mathcal{N}_{\rm abs}^{\rm int}(N_{\rm HI}) = \sum_{i = 1}^{n}\int_{z_{\rm min}}^{z_{\rm max}}\int_{0}^{N_{\rm HI}} f(N_{\rm HI}', z)\,C_i(N_{\rm HI}', z)     \\w_{i}(z+\Delta{z}_{\rm asc})\,\mathrm{d}N_{\rm HI}'\,\mathrm{d}{X(z)},
\end{multline}
where $f(N_{\rm HI}, z)$ is the $N_{\rm HI}$ frequency distribution function, equal to the number of systems with a column density between $N_{\rm HI}$ and $N_{\rm HI} + \mathrm{d}N_{\rm HI}$ per unit column density per unit comoving absorption path length. 

There have been several measurements of $f(N_{\rm HI}, z)$, using 21-cm emission line surveys in the local Universe (e.g. \citealt{Zwaan:2005a, Braun:2012}) and DLAs at cosmological distances (e.g. \citealt{Noterdaeme:2012, Neeleman:2016, Rao:2017, Bird:2017}). Searches for DLAs at UV-wavelengths using \textit{HST} by \cite{Neeleman:2016} and \cite{Rao:2017} are most closely matched in redshift to FLASH, but the sample sizes are not sufficient to provide precise measurements of $f(N_{\rm HI}, z)$ across the full range of DLA column densities. We therefore interpolate between the results of \cite{Zwaan:2005a} from a 21-cm emission line survey at $z = 0$, and \cite{Bird:2017}, who obtained the most precise measurement yet at $z = 2$ using a catalogue generated by a Gaussian Process (GP) method from the  Sloan Digital Sky Survey III Data Release 12 (\citealt{Garnett:2017}). 


\subsection{Number of associated 21-cm absorbers}\label{section:no_associated_absorbers}

We calculate the expected number of associated 21-cm absorbers by evaluating the following sum over $n$ sight lines
\begin{equation}\label{equation:associated_absorbers}
	\mathcal{N}_{\rm abs}^{\rm asc} = \sum_{i = 1}^{n}\int_{z_{\rm min}}^{z_{\rm max}}\,\lambda_{\rm asc}\,\mathcal{\hat{N}}_{\rm src}(z)\,C_i(\tau, \Delta{v}_{\rm FWHM}, z)\,\mathrm{d}{z},
\end{equation}
where $\lambda_{\rm asc}$ is the detection rate of associated absorbers, $\mathcal{\hat{N}}_{\rm src}(z)$ is the normalised distribution of source redshifts given by \autoref{equation:normalised_source_redshifts}, and $C_i(\tau, \Delta{v}_{\rm FWHM}, z)$ is the completeness for absorption lines of peak optical depth $\tau$ and width $\Delta{v}_{\rm FWHM}$ at redshift $z$. 


\subsection{Measuring $\Omega_{\rm HI}$ from absorption}\label{section:omega_HI}

The cosmological mass density in \mbox{H\,{\sc i}}, as a fraction of the critical density $\rho_{\rm c}$, is given by 
\begin{equation}
\Omega_{\rm HI}(z) \equiv \frac{m_{\rm H}\,H_{0}}{c\,\rho_{\rm c}}\,\int_{0}^{\infty}{N_{\rm HI}}\,f(N_{\rm HI}, z)\,\mathrm{d}N_{\rm HI},
\end{equation}
where $f(N_{\rm HI}, z)$ is the column density distribution function as defined above, $H_{0}$ is the Hubble Constant and $m_{\rm H}$ is the mass of the hydrogen atom. For a discrete sample of $\mathcal{N}_{\rm abs}^{\rm int}$ absorbers, an average measurement of $\Omega_{\rm HI}$ can be obtained by evaluating the following sum
\begin{equation}\label{equation:omega_HI}
	\langle{\Omega_{\rm HI}}\rangle_{\rm abs} = \frac{m_{\rm H}\,H_{0}}{c\,\rho_{\rm c}}\,\sum_{i = 1}^{\mathcal{N}_{\rm abs}^{\rm int}}\,\frac{N_{{\rm HI}, i}}{\Delta{X(N_{{\rm HI}, i})}},
\end{equation}
with standard deviation
\begin{equation}
	\sigma(\Omega_{\rm HI})_{\rm abs} = \frac{m_{\rm H}\,H_{0}}{c\,\rho_{\rm c}}\,\sqrt{\sum_{i = 1}^{\mathcal{N}_{\rm abs}^{\rm int}}\,\left(\frac{N_{{\rm HI}, i}}{\Delta{X(N_{{\rm HI}, i})}}\right)^{2}},
\end{equation}
where $\Delta{X}$ is the total comoving absorption path length sensitive to a given column density, as defined in \autoref{equation:total_absorption_path}. 

By taking the continuous limit of the above, we can determine the expected value measured from FLASH  as follows
\begin{equation}
	\langle{\Omega_{\rm HI}}\rangle_{\rm abs} = \frac{m_{\rm H}\,H_{0}}{c\,\rho_{\rm c}} \, \int_{0}^{\infty} \, \frac{N_{\rm HI}}{\Delta{X}(N_{\rm HI})}\,\frac{\mathrm{d}\mathcal{N}_{\rm abs}^{\rm int}}{\mathrm{d}N_{\rm HI}}\,\mathrm{d}N_{\rm HI},
\end{equation}
with standard deviation
\begin{equation}
	\sigma(\Omega_{\rm HI})_{\rm abs} = \frac{m_{\rm H}\,H_{0}}{c\,\rho_{\rm c}} \, \sqrt{\int_{0}^{\infty} \, \left(\frac{N_{\rm HI}}{\Delta{X}(N_{\rm HI})}\right)^{2}\,\frac{\mathrm{d}\mathcal{N}_{\rm abs}^{\rm int}}{\mathrm{d}N_{\rm HI}}\,\mathrm{d}N_{\rm HI}},
\end{equation}
where $\mathcal{N}_{\rm abs}^{\rm int}$ is the expected number of intervening 21-cm absorbers given by \autoref{equation:intervening_absorbers}. Since the survey is incomplete for low column densities, this measurement is a lower limit to the total \mbox{H\,{\sc i}} mass density.

\end{appendix}


\bibliographystyle{pasa-mnras}
\bibliography{flash_survey}

\end{document}